\documentclass[final]{article}
\usepackage{amsfonts,amssymb,epsfig,subfigure,enumerate,anysize}

\marginsize{1in}{1in}{1.5cm}{2.5cm}

\input{Qcircuit.tex}
\usepackage{graphicx,color}
\usepackage[full]{complexity}

\renewcommand{\Qcircuit}[1][0em]{\xymatrix @*=<#1>}

\newtheorem{theorem}{Theorem}[section]

\newtheorem{lemma}[theorem]{Lemma}

\newtheorem{definition}[theorem]{Definition}
\definecolor{gray}{gray}{0.7}
\newcommand{\qed}{\nobreak \ifvmode \relax \else
      \ifdim\lastskip<1.5em \hskip-\lastskip
      \hskip1.5em plus0em minus0.5em \fi \nobreak
      \vrule height0.75em width0.5em depth0.25em\fi}

\usepackage{algorithm}% http://ctan.org/pkg/algorithm
\usepackage{algorithmic}
\usepackage{amsmath}
\usepackage{tikz}
\usetikzlibrary{calc}
\usetikzlibrary{shapes}

\makeatletter
\newdimen\@myBoxHeight%
\newdimen\@myBoxDepth%
\newdimen\@myBoxWidth%
\newdimen\@myBoxSize%
\newcommand{\SquareBox}[2][]{%
    \settoheight{\@myBoxHeight}{#2}% Record height of box
    \settodepth{\@myBoxDepth}{#2}% Record depth of box
    \settowidth{\@myBoxWidth}{#2}% Record width of box
    \pgfmathsetlength{\@myBoxSize}{max(\@myBoxWidth,(\@myBoxHeight+\@myBoxDepth))}%
    \tikz \node [shape=rectangle, shape aspect=1,draw=black,inner sep=2\pgflinewidth, minimum size=\@myBoxSize,#1] {#2};%
}%
\makeatother

\title{Reversible Logic Synthesis by Quantum Rotation Gates}
\author{Afshin Abdollahi\footnote{Current address: Knight Capital Americas LLC, 545 Washington Blvd, Jersey City, NJ 07310.}, Mehdi Saeedi\footnote{Corresponding author: msaeedi@usc.edu.}, Massoud Pedram\\
\small Department of Electrical Engineering, University of Southern California\\
\small Los Angeles, CA 90089-2562\\
}

\begin{document}
\date{}
\maketitle
%\setlength{\textheight}{8.0truein}    %FOR 2ND PAGE ONWARDS

%\centerline{\bf
%%%%%%%%%%%%%%%%%%%%%%
%%Put in titiles here
%%%%%%%%%%%%%%%%%%%%%%
%REVERSIBLE LOGIC SYNTHESIS BY QUANTUM ROTATION GATES}
%\vspace*{0.37truein}
%\centerline{\footnotesize AFSHIN ABDOLLAHI\footnote{Current address: Knight Capital Americas LLC, 545 Washington Blvd, Jersey City, NJ 07310.}, MEHDI SAEEDI\footnote{Corresponding author: msaeedi@usc.edu.}, MASSOUD PEDRAM}
%\vspace*{2pt}
%\centerline{\footnotesize\it Department of Electrical Engineering, University of Southern California}
%\centerline{\footnotesize\it Los Angeles, CA 90089-2562}
%\vspace*{0.225TRUEIN}

%\publisher{November x, 2012} {}

%\vspace*{0.21truein}
\renewcommand{\thefootnote}{\arabic{footnote}}
\begin{abstract}
A rotation-based synthesis framework for reversible logic is proposed. We develop a canonical representation based on binary decision diagrams and introduce operators to manipulate the developed representation model. Furthermore, a recursive functional bi-decomposition approach is proposed to automatically synthesize a given function. While Boolean reversible logic is particularly addressed, our framework constructs intermediate quantum states that may be in superposition, hence we combine techniques from reversible Boolean logic and quantum computation. The proposed approach results in quadratic gate count for multiple-control Toffoli gates without ancillae, linear depth for quantum carry-ripple adder, and quasilinear size for quantum multiplexer.
\end{abstract}

\section{Introduction} \label{sec:intro}
The appeal for research on quantum information processing \cite{NeilsenChuang} is due to three major reasons. {\bf (1)} Working with information encoded at the atomic scale such as ions and even elementary particles such as photons is a scientific advance. {\bf (2)} Direct manipulation of quantum information may create new capabilities such as ultra-precise measurement \cite{Giovannetti2011}, telemetry, and quantum lithography \cite{kothe2011efficiency}, and computational simulation of quantum-me\-cha\-nical phenomena \cite{Lanyon2011}. {\bf (3)} Some time-exponential computational tasks with non-quantum input and output have efficient quantum algorithms \cite{NeilsenChuang}. Particularly, most quantum circuits achieve a quantum speed-up over conventional algorithms \cite{Aaronson2011}. However, useful applications remain limited.

Recent advances in fault-tolerant quantum computing decrease per-gate error rates below the threshold estimate \cite{Brown2011} promising larger quantum computing systems. To be able to do efficient quantum computation, one needs to have an efficient set of computer-aided design tools in addition to the ability of working with favorable complexity class and controlling quantum mechanical systems with a high fidelity and long coherence times. This is comparable with the classical domain where a Turing machine, a high clock speed and no errors in switching were not adequate to design fast modern computers.

Quantum circuit design with algorithmic techniques and CAD tools has been followed by several researchers. The proposed methods either addressed permutation matrices \cite{SaeediM2011} or unitary matrices, e.g., \cite{Shende06}. Permutation matrices and reversible circuits are an important class of computations that should be efficiently performed for the purpose of efficient quantum computation. Indeed, Boolean reversible circuits have attracted attention as components in several quantum algorithms including Shor's quantum factoring \cite{MarkovQIC2012,Markov2013} and stabilizer circuits \cite{garcia2012efficient}.

In this paper, a canonical decision diagram-based representation is presented with novel techniques for synthesis of circuits with binary inputs. This work may be considered along with the work done for the synthesis of reversible circuits \cite{SaeediM2011}. However, we work with rotation-based gates which allow computing a Boolean function by leaving the Boolean domain \cite{MaslovTCAD11}. Therefore, this approach may be viewed as a step to explore synthesis of reversible functions by gates other than generalized Toffoli and Fredkin gates. {We show that applying the proposed approach improves (1) circuit size for multiple-control Toffoli gates from exponential in \cite[Lemma 7.1]{Barenco95} to polynomial and from $48n^2+O(n)$ \cite[Lemma 7.6]{Barenco95} to $2n^2+O(n)$, (2) circuit depth for quantum carry-ripple adders by a constant factor compared to \cite{Cuccaro2004}, and (3) circuit size for quantum multiplexers from $O(n^2)$ to $O(n\log^2 n)$.}

The reminder of this paper is organized as follows. In Section \ref{sec:basic_concepts}, we touch upon necessary background in reversible and quantum circuits. {Readers familiar with quantum circuits may ignore this section.} Section \ref{sec:prior_work} summarizes the previous work on quantum and reversible circuit synthesis. In Section \ref{sec:proposed}, the proposed rotation-based technique is described. In Section \ref{sec:extension}, we provide an extension of the proposed synthesis algorithm to handle a more general logic functions, i.e., functions with binary inputs and arbitrary outputs. Synthesis of several function families are discussed in Section \ref{sec:ex}, and finally Section \ref{sec:conc} concludes the paper. {A partial version of this paper was presented in \cite{Afshin_06}.}

\section{Basic Concepts} \label{sec:basic_concepts}
A quantum bit, \emph{qubit}, can be realized by a physical system such as a photon, an electron or an ion. In this paper, we treat a qubit as a mathematical object which represents a quantum state with two basic states $\ket{0}$ and $\ket{1}$. A qubit can get any linear combination of its basic states, called \emph{superposition}, as $|\psi\rangle = \alpha|0\rangle+\beta|1\rangle$ where $\alpha$ and $\beta$ are complex numbers and $|\alpha|^2+|\beta|^2= 1$.

Although a qubit can get any linear combination of its basic states, when a qubit is \emph{measured}, its state collapses into the basis $|0\rangle$ and $|1\rangle$ with the probability of $|\alpha|^2$ and $|\beta|^2$, respectively. It is also common to denote the state of a single qubit by a $2 \times 1$ vector as $[\begin{array}{*{20}c}   \alpha  & \beta   \\ \end{array}]^T$ in Hilbert space $H$ where superscript $T$ stands for the transpose of a vector. A quantum system which contains $n$ qubits is often called a \emph{quantum register} of size $n$. Accordingly, an $n$-qubit quantum register can be described by an element $|\psi \rangle = |\psi_1 \rangle \otimes |\psi_2 \rangle \otimes \ldots \otimes |\psi_n \rangle$ (simply $|\psi_1\psi_2\cdots\psi_n \rangle$) in the tensor product Hilbert space $H = H_1 \otimes H_2 \otimes \dots \otimes H_n$.

An $n$-qubit \emph{quantum gate} performs a specific $2^n\times2^n$ unitary operation on selected $n$ qubits. A matrix $U$ is unitary if $UU^\dag=I$ where $U^\dag$ is the conjugate transpose of $U$ and $I$ is the identity matrix. The unitary matrix implemented by several gates acting on different qubits independently can be calculated by the tensor product of their matrices. Two or more quantum gates can be cascaded to construct a \emph{quantum circuit}. For a set of $k$ gates $g_1, g_2, \cdots, g_k$ cascaded in a quantum circuit $C$ in sequence, the matrix of $C$ can be calculated as $M_k M_{k-1} \cdots M_1$ where $M_i$ is the matrix of the $i^{th}$ gate ($1\leq i \leq k$). For a quantum circuit with unitary matrix $U$ and input vector $\psi_1$, the output vector is $\psi_2 = U \psi_1$.

Various quantum gates with different functionalities have been introduced. The $\theta$-rotation gates ($0 \leq \theta \leq 2\pi$) around the $x$, $y$ and $z$ axes acting on one qubit are defined as Eq. (\ref{eq:rotation}). The single-qubit NOT gate is described by the matrix $X$ in Eq. (\ref{eq:CNOT_H}). The CNOT (controlled NOT) acts on two qubits (control and target) is described by the matrix representation shown in Eq. (\ref{eq:CNOT_H}). The Hadamard gate, $H$, has the matrix representation shown in Eq. (\ref{eq:CNOT_H}).

\begin{small}
\begin{equation} \label {eq:rotation}
     R_x (\theta ) = \left( {\begin{array}{*{20}c}
       {\cos \frac{\theta }{2}} & {-i\sin \frac{\theta }{2}}  \\
       {-i\sin \frac{\theta }{2}} & {\cos \frac{\theta }{2}}  \\
    \end{array}} \right) ,
     R_y (\theta ) = \left( {\begin{array}{*{20}c}
       {\cos {\textstyle{\frac{\theta}{2}}}} & {-\sin {\textstyle{\frac{\theta}{2}}}}  \\
       {  \sin {\textstyle{\frac{\theta}{2}}}} & {\cos {\textstyle{\frac{\theta}{2}}}}  \\
    \end{array}} \right),
    R_z (\theta) = \left( {\begin{array}{*{20}c}
   {e{\textstyle{\frac{ - i\theta }{2}}}} & 0  \\
   0 & {e{\textstyle{\frac{i\theta }{2}}}}  \\
    \end{array}} \right)
\end{equation}
\end{small}

\begin{small}
\begin{equation} \label {eq:CNOT_H}
X = \left( {\begin{array}{*{20}c}
   0 & 1  \\
   1 & 0  \\
\end{array}} \right),
\rm{CNOT}=\left( {\begin{array}{*{20}c}
   1 & 0 & 0 & 0  \\
   0 & 1 & 0 & 0  \\
   0 & 0 & 0 & 1  \\
   0 & 0 & 1 & 0  \\
\end{array}} \right),
H=\frac{1}{{\sqrt 2 }}\left( {\begin{array}{*{20}c}
   1 & 1  \\
   1 & { - 1}  \\
\end{array}} \right)
\end{equation}
\end{small}

\begin{figure}[tb]
\scriptsize
\scalebox{0.9}{
\input{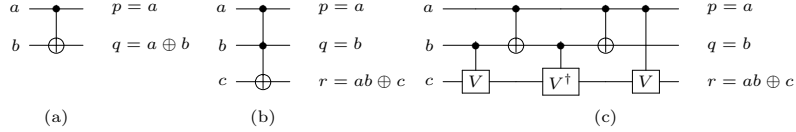}
}
\centering
\vspace{2mm}
\caption{\label{Fig:CTGates-org} CNOT (a) and Toffoli (b) gates. Decomposition of a Toffoli gate into 2-qubit gates (c) where $V=(1-i)(I+iX)/2$ \cite{Barenco95}.}
\end{figure}

Given any unitary $U$ over $m$ qubits $\ket{x_1 x_2  \cdots \,x_m}$, a controlled-$U$ gate with $k$ control qubits $\ket{y_1 y_2  \cdots \,y_k}$ may be defined as an $(m+k)$-qubit gate that applies $U$ on $\ket{x_1 x_2  \cdots \,x_m }$ iff $\ket{y_1 y_2  \cdots \,y_k}$=$\ket{11\cdots1}$. For example, CNOT is the controlled-NOT with a single control, Toffoli is a NOT gate with two controls, and $CR_x(\theta)$ is a $R_x(\theta)$ gate with a single control. Similarly, a multiple-control Toffoli gate C$^k$NOT is a NOT gate with $k$ controls. Fig. \ref{Fig:CTGates-org} shows CNOT and Toffoli gates. For a circuit implementing a unitary $U$, it is possible to implement a circuit for the controlled-$U$ operation by replacing every gate by a controlled gate. In circuit diagrams, $\bullet$ is used for conditioning on the qubit being set to value one.

\section{Previous Work} \label{sec:prior_work}
Synthesis of 0-1 unitary matrices, also called \emph{permutation}, has been followed by several researchers during the recent years. Here, we review the recent approaches with favorable results. More information can be found in \cite{SaeediM2011}. \emph{Transformation-based} methods \cite{MaslovTODAES07} iteratively select a gate to make a given function more similar to the identity function. These methods construct compact circuits mainly for permutations with repeating patterns in output codewords. \emph{Search-based} methods \cite{DonaldJETC08} explore a search tree to find a realization. These methods are highly useful if the number of circuit lines and the number of gates in the final circuit are small. \emph{Cycle-based} methods \cite{SaeediJETC10} decompose a given permutation into a set of disjoint (often small) \emph{cycles} and synthesize individual cycles separately. These methods are mainly efficient for permutations without repeating output patterns. \emph{BDD-based} methods \cite{Wille_09} use binary decision diagrams to improve sharing between controls of reversible gates. These techniques scale better than others. However, they require a large number of ancilla qubits.

Quantum-logic synthesis deals with general unitary matrices and is more challenging than reversible-logic synthesis. Synthesis of an arbitrary unitary matrix from a universal set of gates including one-qubit operations and CNOTs has a rich history. Barenco et al. in 1995 \cite{Barenco95} showed that the number of CNOT gates required to implement an arbitrary unitary matrix over $n$ qubits was $O(n^34^n)$. As of 2012, the most compact circuit constructions use $\frac{23}{48} 4^n  - \frac {3}{2} 2^n  + \frac{4}{3}$ CNOTs \cite{Shende06,Mottonen06} and $\frac{1}{2} 4^n  + \frac{1}{2} 2^n  - n - 1$ one-qubit gates \cite{Bergholm:2004}. The sharpest lower bound on the number of CNOT gates is $\left\lceil \frac{1}{4}(4^n  - 3n - 1) \right\rceil$ \cite{shende-2004}. Different trade-offs between the number of one-qubit gates and CNOTs are explored in \cite{SaeediQIC11}.

\section{Rotation-Based Synthesis of Boolean Functions} \label{sec:proposed}
\begin{figure}[tb]
\scriptsize
\scalebox{0.8}{
\input{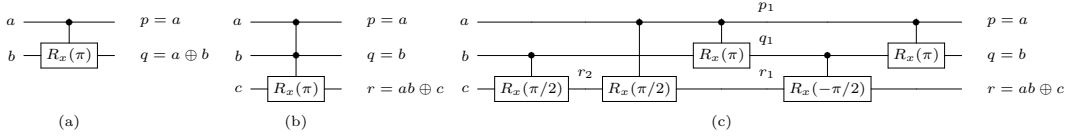}
}
\centering
\vspace{2mm}
\caption{\label{Fig:CTGates} New definitions for CNOT (a) and Toffoli (b) gates using controlled rotation gates. Decomposition of a Toffoli gate into 5 2-qubit controlled-rotation gates (c).}
\end{figure}

In this section, we address the problem of automatically synthesizing a given Boolean function $f$ by using rotation and controlled-rotation gates around the $x$ axis. In this paper, we change the basis states as $\hat 0 = \left[ {\begin{array}{*{20}c}  1 & 0  \\ \end{array}} \right]^T$ and $\hat 1=R_x(\pi)\hat 0=  \left[ {\begin{array}{*{20}c} 0 & -i  \\ \end{array}} \right]^T$. With this definition of $\hat 0$ and $\hat 1$, the basis states remain orthogonal.
Also, inversion (i.e., the NOT gate) from one basis state to the other is simply obtained by a $R_x(\pi)$ gate.\footnote{While we used $\hat 0$ and $\hat 1$ as the basis states, the presented algorithm can be easily modified to be applicable to quantum functions described in terms of $\ket{0}$ and $\ket{1}$. An alternate solution is to define the following operators and use $M$ to transform the $\ket{0}$ and $\ket{1}$ states to $\hat 0$ and $\hat 1$ states and operator $M^{-1}$ to perform the reverse transformation. Hence to compute in $\ket{0}$ and $\ket{1}$ basis, one needs to apply $M$ and $M^{-1}$ single-qubit operators before and after the computation done in the $\hat 0$ and $\hat 1$ basis, respectively. Notice that $M$ and $M^{-1}$ are rotations around the $z$ axis.
$$
M = \left[ {\begin{array}{*{20}c} 1 & 0  \\ 0 & { - i}  \\ \end{array}} \right],
M^{-1} = \left[ {\begin{array}{*{20}c} 1 & 0   \\ 0 & i  \\ \end{array}} \right]
$$
}
Subsequently, the CNOT gate can be described by using the $CR_x(\pi)$ operator shown in Fig. \ref{Fig:CTGates}(a). In addition, the Toffoli gate may be described by using the $C^2R_x(\pi)$ operator illustrated in Fig. \ref{Fig:CTGates}(b). Toffoli gate can be implemented using 5 controlled-rotation operators as demonstrated in Fig. \ref{Fig:CTGates}(c). Recall that a 3-qubit Toffoli gate needs 5 2-qubit gates if $\ket{0}$ and $\ket{1}$ are used as the basis states (Fig. \ref{Fig:CTGates-org}(c)).

For a 2-qubit $CR_x(\theta)$ gate with a control qubit $a$ and a target qubit $b$, the first output is equal to $a$. However, the second output depends on both the control line $a$ and the target line $b$. We use the notation $aR_x(\theta)b$ to describe the second output. Furthermore, we write $R_x(\theta)b$ to unconditionally apply a single-qubit $R_x(\theta)$ to the qubit $b$. Additionally, one can show that for binary variables $a, b, c$ we have $aR_x(\theta_1)[aR_x(\theta_2)b]=aR_x(\theta_1+\theta_2)b$, $aR_x (\theta _1 )\left[ {bR_x (\theta _2 )c} \right] = bR_x (\theta _2 )\left[ {aR_x (\theta _1 )c} \right]$, $a R_x(\pi) \hat 1=\sim a$ ($\sim$ is used for negation), and $a R_x(\pi) \hat 0=a$.

\begin{definition}
\normalfont
$\hat 0$ and all variables are in the \emph{rotation-based factored} (factored in short) form. If $h$ and $g$ are in the factored form, then $R_x(\theta)h$ and $gR_x(\theta)h$ are in the factored form too.
\end{definition}

In a quantum circuit synthesized with $R_x(\theta)$ and $CR_x(\theta)$ operators, all outputs and intermediate signals in the given circuit can be described in the factored form. For example, the output $r$ in Fig. \ref{Fig:CTGates}(c) can be described as $r = \left[ {aR_x (\pi )b} \right]R_x ( - \pi /2)\left[ {aR_x (\pi /2)\left[ {bR_x (\pi /2)c} \right]} \right]$.

\begin{definition}
\normalfont
$\hat 0$ and all variables are \emph{rotation-based cascade} (cascade in short) expressions. If $h$ is a cascade expression and $v$ is a variable $\not \in h$, then $R_x(\theta)h$ and $vR_x(\theta)h$ are cascade expressions too ($\forall \theta$).
\end{definition}

A cascade expression can be expressed as ${R_x (\theta _0 )\left[ {v_1 R_x (\theta _1 )\left[ {v_2 R_x (\theta _2 ){\rm  }\cdots{\rm}\left[ {v_n R_x (\theta _n )\hat 0} \right]{\rm  }} \right]{\rm }} \right]}$.
The problem of realizing a function with $R_x (\theta)$ and $CR_x (\theta)$ operators is equivalent to finding a cascade expression for the function. To do this, we first introduce a graph-based data structure in the form of a decision diagram for representing functions.

\subsection{A Rotation-Based Data Structure} \label{sec:RbDD}
The concept of binary decision diagram (BDD) was first proposed by Lee \cite{Lee} and later developed by Akers \cite{Akers} and then by Bryant \cite{Bryant:1986}, who introduced Reduced Ordered BDD (ROBDD) and proved its canonicity property. Bryant also provided a set of operators for manipulating ROBDDs. In this paper, we omit the prefix RO. BDD has been extensively used in classical logic synthesis. Furthermore, several variants of BDD were also proposed for logic synthesis \cite{Wille_09}, verification \cite{Viamontes:2007,Yamashita08,Wang:2008} and simulation \cite{MillerISMVL06,Viamontes09} of reversible and quantum circuits. In this section, we describe a new decision diagram for the representation of functions based on rotation operators. Next, we use it to propose a synthesis framework for logic synthesis with rotation gates.

\begin{definition}
\normalfont
A Rotation-based Decision Diagram (RbDD) is a directed acyclic graph with three types of nodes: a single terminal node with value $\hat 0$, a weighted root node, and a set of non-terminal (internal) nodes. Each internal node represents a function and is associated with a binary decision variable with two outgoing edges: a weighted $\hat 1$-edge (solid line) leading to another node, the $\hat 1$-child, and a non-weighted  $\hat 0$-edge (dashed line) leading to another node, the $\hat 0$-child. The weights of the root node and  $\hat 1$-edges are in the form of $R_x(\theta)$ matrices. We assume that $-\pi < \theta \leq \pi$. When a weight (either for an edge or the root node) is the identity matrix (i.e., $R_x(0)=I$), it is not shown in the diagram.
\end{definition}

\begin{figure}[tb]
    \centering
    \fbox{
    \subfigure[\label{Fig:RbDD} ]{
          \includegraphics[height=2.5cm]{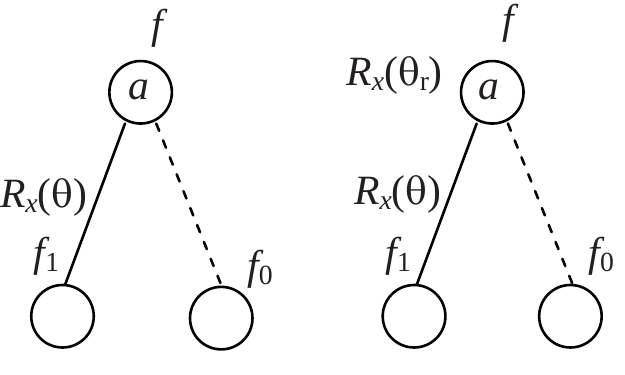}
    }
    }
    \fbox{
    \subfigure[\label{Fig:linear} ]{
          \includegraphics[height=2.5cm]{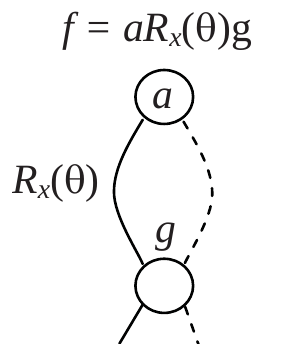}
    }
    \subfigure[\label{Fig:linearCirc} ]{
          \includegraphics[height=1.25cm]{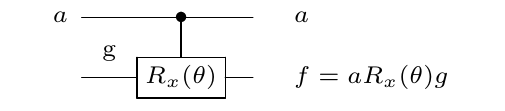}
    }
    }
    \vspace{2mm}
    \caption{(a) Internal structure of a rotation-based decision diagram (RbDD) without and with a weighted root. (b) For a node $f$, if the $\hat 0$-child and the $\hat 1$-child are the same node $g$, $f$ can be directly realized by a $R_x(\theta)$ operator as shown in (c).}
\end{figure}

The left RbDD in Fig. \ref{Fig:RbDD} shows an internal node $f$ with decision variable $a$, the corresponding $\hat 0$  and $\hat 1$ edges, and child nodes $f_0$ and $f_1$. The relation between the RbDD nodes in this figure is as follows. If $a = \hat 1$, then $f = R_x (\theta )f_1$ else $f=f_0$. In addition, if $f$ is a weighted root node as shown in the right RbDD in Fig. \ref{Fig:RbDD}, then for $a = \hat 1$ we have $f = R_x (\theta _r )R_x (\theta )f_1  = R_x (\theta _r  + \theta )f_1 $; otherwise $f = R_x (\theta _r )f_0$. Similar to BDDs, in RbDDs isomorphic sub-graphs which are nodes with the same functions are merged. Additionally, if the $\hat 0$-child and the $\hat 1$-child of a node are the same and the weight of $\hat 1$-edge is $R_x(0)=I$, then that node is eliminated. Using these two reduction rules and a given total ordering $\prec$ on input variables, one can uniquely construct the RbDD of a given function. {Notably, a decision diagram called DDMF was proposed in \cite{Yamashita08}, where each edge can represent any unitary matrix including rotation operators. DDMF was used for verification of quantum circuits.}

\begin{figure}[tb]
    \centering
    \fbox{
    \subfigure[\label{Fig:TQDD1} ]{
        \includegraphics[width=7.8cm]{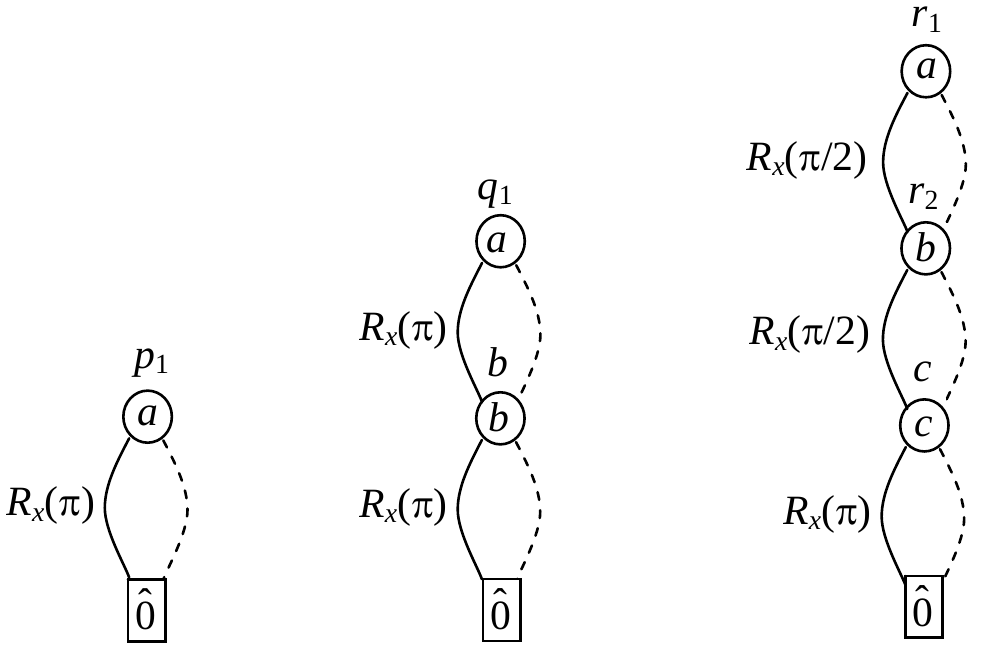}
    }
    \subfigure[\label{Fig:TQDD2} ]{
          \includegraphics[height=1.55cm]{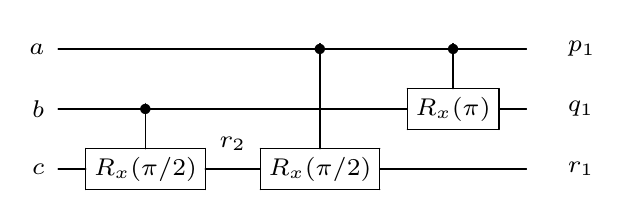}
    }
    }
    \vspace{2mm}
    \caption{RbDDs for intermediate signals of a 3-input Toffoli gate shown in Fig. \ref{Fig:CTGates}(c), redrawn in (b). In this figure, we have $ q_1 = a Rx(\pi) b$, $r_1 = a R_x(\pi/2) r_2$, and $r_2 = b R_x(\pi/2) c$.
}
\end{figure}

For a given function $f$ with $n$ binary variables $v_1, v_2, \cdots, v_n$, each value assignment to $v_1, v_2, \cdots, v_n$ corresponds to a path from the root to the terminal node in the RbDD of $f$. Assuming the variable ordering $v_1<v_2< \cdots <v_n$, the corresponding path can be identified by a top-down traversal of the RbDD starting from the root node. For each node visited during the traversal, we select the edge corresponding to the value of its decision variable $v_i$. Denote the weight of the root node by $w_0$ and the weight of the selected edges by $w_1, w_2, \cdots, w_{n-1}$. We have $f(v_1 ,v_2 ,\cdots,v_n ) = w_0 w_1 \cdots w_{n - 1} \hat 0 = w_0 w_1 \cdots w_{n - 1} \left[ {\begin{array}{*{20}c}  1  &   0  \end{array}} \right]^T$. If a $\hat 0$-edge is selected for variable $v_i$ (i.e., if $v_i= \hat 0$), we have $w_i=I$. Note that when the $\hat 0$-child and the $\hat 1$-child of a node $f$ are the same node $g$, then that node can be directly realized by a $R_x(\theta)$ operator, as $f = aR_x (\theta )g$ demonstrated in Fig. \ref{Fig:linear} and Fig. \ref{Fig:linearCirc}, in terms of its child. Fig. \ref{Fig:TQDD1} shows the RbDDs of functions $p_1$, $q_1$ and $r_1$ in Fig. \ref{Fig:CTGates}(c) (reproduced in Fig. \ref{Fig:TQDD2}). Every RbDD with a chain structure such as the ones shown in Fig. \ref{Fig:TQDD1} is associated with a cascade expression and can be realized with rotation and controlled-rotation operators.

\begin{figure}[tb]
    \centering
    \fbox{
    {
    \subfigure[\label{Fig:Rule1} ]{
          \includegraphics[height=2.25cm]{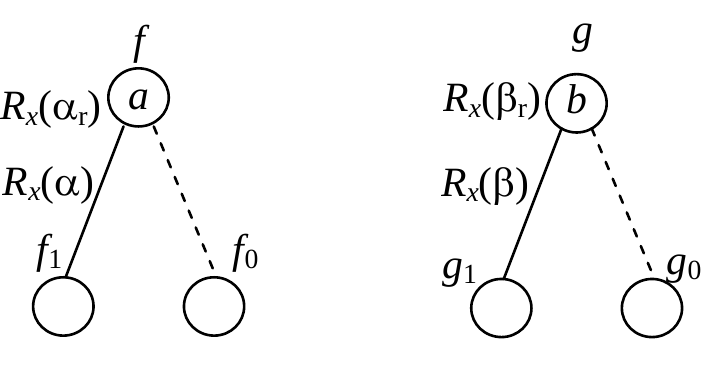}
    }
    }
    }
    \fbox{
    {
    \subfigure[\label{Fig:Rule2} ]{
          \includegraphics[height=2.25cm]{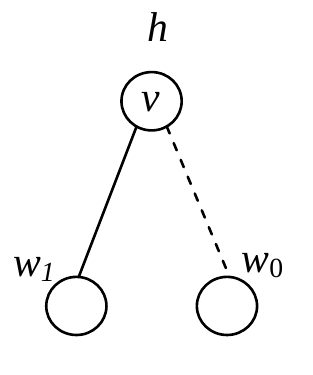}
    }
    }
    }
    \fbox{
    {
    \subfigure[\label{Fig:qapplyW} ]{
        \includegraphics[height=2.25cm]{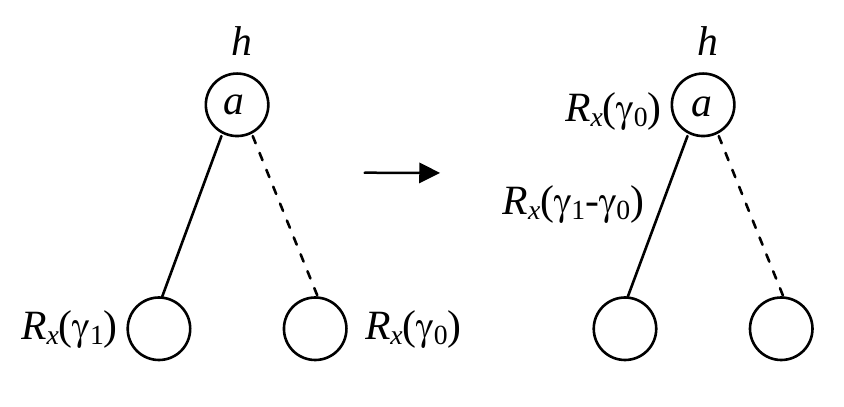}
    }
    }
    }
    \vspace{2mm}
    \caption{\label{Fig:QDDRules} (a) Operands for operation $h =  f R_x(\gamma) g$. (b) The result of \texttt{apply} operator which adds a new node to the resulting RbDD $h$ by using one of the three rules: if $a<b$ (Rule 1), $v=a$, $w_1= [R_x(\alpha_r+\alpha)f_1] R_x(\gamma) g$, $w_0= [R_x(\alpha_r)f_0] R_x(\gamma) g$. If $b<a$ (Rule 2), $v=b$, $w_1=f R_x(\gamma) [R_x(\beta_r+\beta)g_1]$, $w_0= f R_x(\gamma) [R_x(\beta_r)g_0]$. If $a=b$ (Rule 3), $v=a=b$, $w_1= [R_x(\alpha_r+\alpha)f_1] R_x(\gamma) [R_x(\beta_r+\beta)g_1]$, $w_0=[Rx(\alpha_r)f_0] R_x(\gamma) [R_x(\beta_r)g_0]$. (c) Weight modification for the \texttt{apply} operator to maintain canonicity of the resulting RbDD.
    }
\end{figure}

%\subsection{Operations on RbDDs} \label{sec:QApply}
Suppose that the RbDD for a function $f$ is given. The RbDD for $h = R_x (\gamma )f$ can be obtained by multiplying the root weight of $f$ by $R_x (\gamma )$. To obtain $h = fR_x (\gamma )g$ for given RbDDs of $f$ and $g$, we use the \texttt{apply} operator.\footnote{In general, for a binary operation \texttt{op} and two BDDs of functions $f$ and $g$, the \texttt{apply} operator computes a BDD for $f\,\rm{\texttt{op}}\,g$ \cite{Bryant:1986}.} In this context, $f$ and $g$ are called RbDD operands of $h$. The \texttt{apply} operator is implemented by a recursive traversal of the two RbDD operands. For  each pair of nodes in $f$ and $g$ visited during the traversal, an internal node is added to the resulting RbDD by utilizing the following rules which depend on the selected variable ordering $\prec$ (also see Fig. \ref{Fig:QDDRules}). We assume that $f$ and $g$ have two general RbDDs shown in Fig. \ref{Fig:Rule1}. The \texttt{apply} operator is recursively called with the terminal conditions $\hat 0R_x (\theta )v = v$ and $\hat 1R_x (\theta )v = R_x (\theta )v$.

\algsetup{indent=2em}
\begin{algorithm}[h!]
%\caption{Rules for apply operator}\label{Alg:qapply}
\begin{algorithmic}[10]
\medskip
\STATE
\begin{itemize}
  \item \textbf{Rule 1 ($a<b$)} The new node for $h$ is $a$. The weights of $\hat 1$-child and $\hat 0$-child are $[R_x(\alpha_r+\alpha)f_1] R_x(\gamma) g$, and $[R_x(\alpha_r)f_0] R_x(\gamma) g$, respectively.
\item \textbf{Rule 2 ($b<a$)} The new node for $h$ is $b$. The weights of $\hat 1$-child and $\hat 0$-child are $f R_x(\gamma) [R_x(\beta_r+\beta)g_1]$, and $f R_x(\gamma) [R_x(\beta_r)g_0]$, respectively.
\item \textbf{Rule 3 ($a=b$)} The new node for $h$ is $a$ (or $b$). The weights of $\hat 1$-child and $\hat 0$-child are $[R_x(\alpha_r+\alpha)f_1] R_x(\gamma) [R_x(\beta_r+\beta)g_1]$, and $[Rx(\alpha_r)f_0] R_x(\gamma) [R_x(\beta_r)g_0]$, respectively.
\end{itemize}
\end{algorithmic}
\end{algorithm}

\begin{figure}
    \centering
    \fbox{
    \subfigure[\label{Fig:qapply} ]{
    \includegraphics[height=5.5cm]{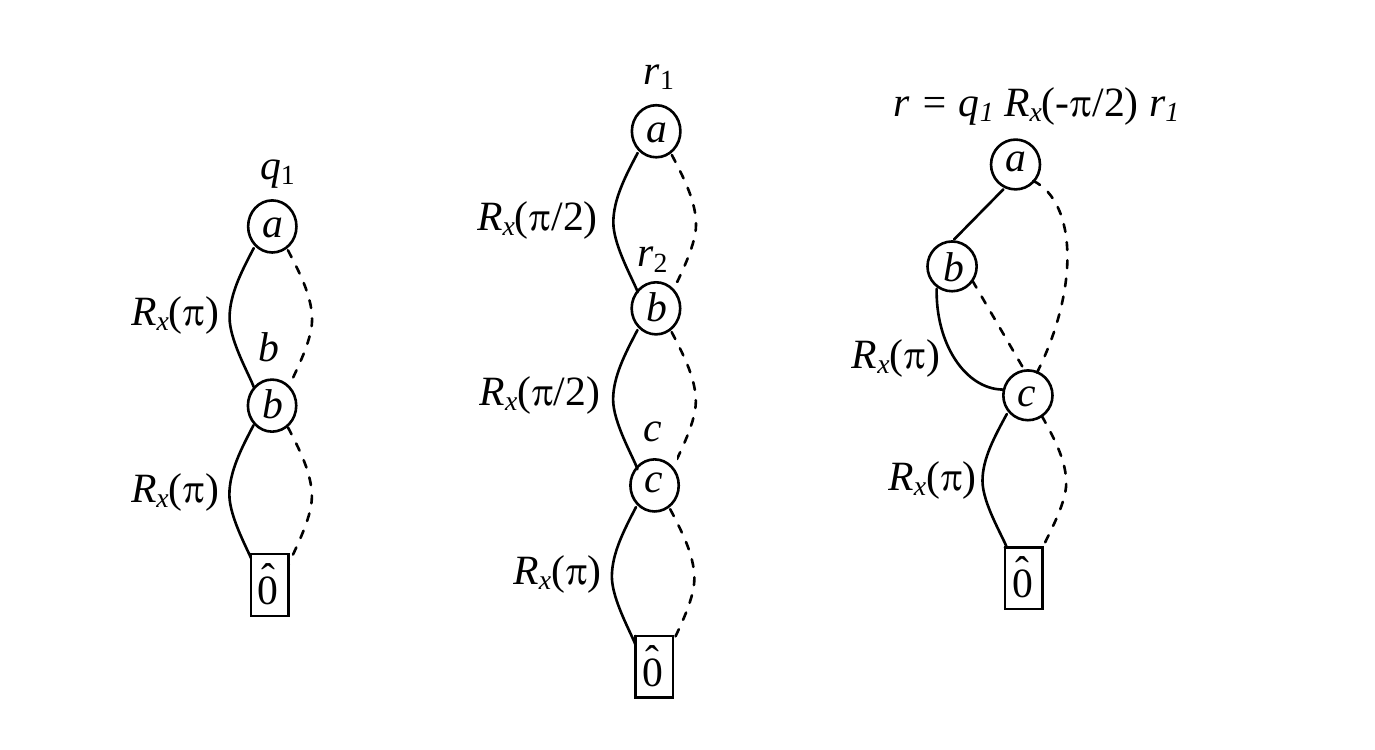}
    }
    }
    \fbox{
    \subfigure[\label{Fig:q_linear}]{
    \includegraphics[height=5.5cm]{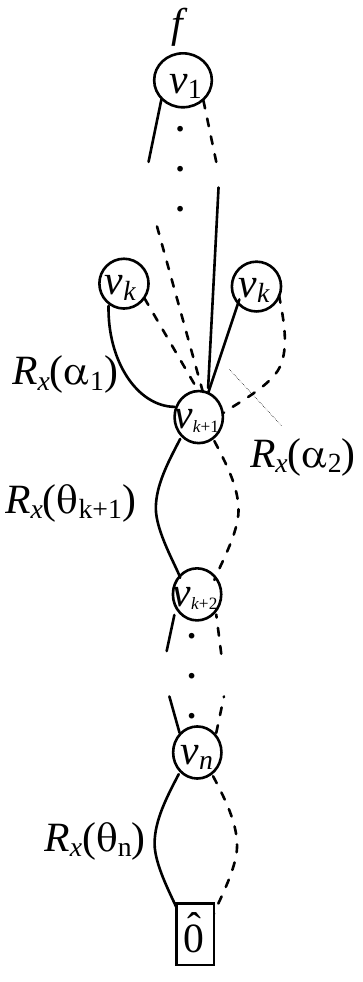}
    }
    }
    \vspace{2mm}
    \caption{(a) An example of performing \texttt{apply} operator on two RbDDs. In the first call of the \texttt{apply} operator, $w_1=[R_x(\pi)b]R_x(-\pi/2)[R_x(\pi/2)r_2]$ and $w_0=bR_x(-\pi/2)r_2$. The final figure is obtained after eliminating redundant nodes and edges. (b) A general RbDD structure with r-linear variables $v_{k+1}, \cdots, v_n$.}
\end{figure}

After recursive computation of the $\hat 1$-child and $\hat 0$-child of $h$, to maintain the canonicity of the resulting RbDD, isomorphic sub-graphs are merged and if the $\hat 0$-child and the $\hat 1$-child of a node are the same and the weight of the $\hat 1$-edge is $R_x(0)=I$, then that node will be eliminated. In addition, to make RbDD of $h$ canonical, the resulting weights for the $\hat 1$-child and the $\hat 0$-child of $h$ should be modified by the method illustrated in Fig. \ref{Fig:qapplyW}. Fig. \ref{Fig:qapply} demonstrates the result of performing \texttt{apply} operator on $q_1$ and $r_1$ in Fig. \ref{Fig:TQDD1}, redrawn in Fig. \ref{Fig:qapply}, to obtain $r=q_1R_x(-\pi/2)r_1$. To construct RbDD for $r$, one needs to initially apply Rule 3 because both $q_1$ and $r_1$ use $a$ as roots. Accordingly, $w_1=[R_x(\pi)b]R_x(-\pi/2)[R_x(\pi/2)r_2]$ and $w_0=bR_x(-\pi/2)r_2$. To continue, consider $w_1$ and note that both $[R_x(\pi)b]$ and $[R_x(\pi/2)r_2]$ use $b$.\footnote{To understand the RbDDs of $[R_x(\pi)b]$ and $[R_x(\pi/2)r_2]$, recall RbDDs of $b$ and $r_2$ in Fig. \ref{Fig:qapply} and use weights $R_x(\pi)$ and $R_x(\pi/2)$ for roots of $b$ and $r_2$, respectively.} As a result, applying Rule 3 leads to $w_{1,1}=[R_x(0)\hat 0] R_x(-\pi/2) [R_x(\pi/2+\pi/2)c]$ and $w_{1,0}=[R_x(\pi)\hat 0] R_x(-\pi/2) [R_x(\pi/2)c]$. On the other hand, applying Rule 3 on $w_0$ leads to $w_{0,1}=[R_x(\pi)\hat0] R_x(-\pi/2) [R_x(\pi/2)c]$ and $w_{0,0}=[R_x(0)\hat0]R_x(-\pi/2)c$. Using terminal conditions results in $w_{1,1}=R_x(\pi)c$, $w_{1,0}=c$, $w_{0,1}=c$, and $w_{0,0}=c$. Since $w_{0,1}=w_{0,0}=c$, we can remove variable $b$ as the $\hat 0$-child of $a$. The final figure in Fig. \ref{Fig:qapply} is obtained after eliminating redundant nodes and edges.\footnote{Note that the commutative property of matrix multiplication for $R_x(\theta)$ matrices is critical for the \texttt{apply} operator. Performing \texttt{apply} as described may not generate the correct result for decision diagrams with non-commutative weights.}

\subsection{Functional Decomposition and r-Linearity}  \label{sec:QDCM}
The problem of realizing a function $f$ using $R_x(\theta)$ and $CR_x(\theta)$ operators is equivalent to finding a rotation-based factored form for $f$, which can be performed by recursive bi-decomposition of $f$.

\begin{definition}
\normalfont
\emph{Rotation-based bi-decomposition} (bi-decomposition in short) of $f$ is defined as finding functions $g$ and $h$ and value $\gamma$ such that $f = gR_x (\gamma )h$.
\end{definition}

We use bi-composition of a given function $f$ to construct $f=gR_x (\gamma )h$. Subsequently, $g$ and $h$ are recursively bi-decomposed, which will eventually result in a factored form of $f$. The bi-decomposition algorithm is based on the notion of r-linearity.

\begin{definition}
\normalfont
For function $f(v_1 ,\cdots,v_{i - 1} ,v_i ,v_{i + 1} ,\cdots,v_n )$, variable $v_i$ is \emph{r-linear} if there exists a rotation value $\theta_i$ such that for every value assignment to $v_1 ,\cdots,v_{i - 1} ,v_{i + 1} ,\cdots,v_n: f_{v_i}  = R_x (\theta _i )f_{{\overline v}_i}$, where $f_{v_i} = f(v_1 ,\cdots,v_{i - 1} ,\hat 1,v_{i + 1} ,\cdots,v_n )$ and $f_{{\overline v}_i}  = f(v_1 ,\cdots,v_{i - 1} ,\hat 0,v_{i + 1} ,\cdots,v_n )$. A variable is \emph{r-nonlinear} if it is not r-linear.
\end{definition}

Now we present a number of key results.

\begin{lemma} \label{lemma:1}
\normalfont
Consider a function $f(v_1 ,v_2 ,\cdots,v_n )$ with variable ordering $v_1  < v_2  < \cdots < v_n$ and assume that $k+1 \leq i \leq n$. Iff each variable $v_i$ is r-linear, then there is only one RbDD node $n_i$ for each r-linear decision variable $v_i$. The weight of the $\hat 1$-edge of $n_i$ is $R_x(\theta_i)$.
\end{lemma}

{\bf Proof.} The proof is by induction on $v_n, v_{n-1}, v_{n-2}, \cdots, v_{k+1}$ starting from $v_n$. \qed

%\vspace{0.5cm}
Let $v_k$ be the lowest indexed r-nonlinear variable after which $v_{k + 1} ,v_{k + 2} ,\cdots,v_n$ are r-linear variables of $f$. From Lemma \ref{lemma:1}, for $k+1 \leq j \leq n$ we have $f_{v_j }  = R_x (\theta _j )f_{\overline v _j }$ where $\theta_j$ is fixed independent of $v_1 ,v_2 ,\cdots,v_{j - 1} ,v_{j + 1} ,\cdots,v_n$ values. As illustrated in Fig. \ref{Fig:q_linear}, every path from the root node of the RbDD to the terminal node will either go through an internal node with decision variable $v_k$ or it will skip any such node and directly go the single RbDD node with decision variable $v_{k+1}$. For the latter case, $f_{v_k }  = R_x (0)f_{\bar v_k }  = f_{\bar v_k }$ and for any former case $f_{v_k }  = R_x (\alpha _i )f_{\bar v_k }$ for some (vs. all) $v_1 ,\cdots,v_{k - 1} ,v_{k + 1} ,\cdots,v_n $. Additionally, the number of different rotation angles (e.g., $\alpha_1, \alpha_2$ in Fig. \ref{Fig:q_linear}) for variable $v_k$ is equal to the number of internal nodes with decision variable $v_k$ in the RbDD.

\begin{definition}\label{def:r-lin}
\normalfont
The \emph{degree of r-nonlinearity} of variable $v_k$, r-deg$(v_i)$, is $m-1$ where $m$ is the number of different rotation angles $\alpha_i$ (including 0 if any) that $f_{v_k }  = R_x (\alpha _i )f_{\bar v_k } $ for some $v_1 ,\cdots,v_{k - 1} ,v_{k + 1} ,\cdots,v_n $. For r-linear variables the degree of r-nonlinearity is zero.
\end{definition}

{As an example, consider the RbDD of $r$ in Fig. \ref{Fig:qapply} and note that r-deg$(b)=1$ as there are two rotation angles (i.e., $0$ and $\pi$) for $b$. Similarly, r-deg$(c)=0$ and $c$ is r-linear.}

\begin{lemma} \label{lemma:2}
\normalfont
Let $m$ denote the number of internal nodes with decision variable $v_k$. If all paths from the root node of the RbDD to the terminal node go through an internal node with decision variable $v_k$, r-deq$(v_k)=m-1$; otherwise r-deg$(v_k)=m$.
\end{lemma}

{\bf Proof.} The proof follows from considering the general structure of RbDDs and the definition of r-nonlinearity.  \qed

\algsetup{indent=2em}
\begin{algorithm}[tb]
\caption{\texttt{factor} ($f$)}\label{Alg:qfact}

\begin{algorithmic}[1]
\medskip
\STATE If all variables are r-linear, then return the corresponding cascade expression for $f$.
\STATE Find the lowest indexed r-nonlinear variable $v_k$ after which $v_{k + 1} ,v_{k + 2} ,\cdots,v_n $  are r-linear.
\STATE Bi-decompose $f$ using $v_k$ as $f = g_1 R_x (\gamma )h$ where $g_1$, $h$, and $\gamma$ are given in Theorem \ref{theorem:1}.
\STATE Return [\texttt{factor($g_1$)}]$R_x(\gamma)$[\texttt{factor($h$)}].
\end{algorithmic}
\end{algorithm}

\begin{theorem} \label{theorem:1}
\normalfont
Consider a function $f(v_1 ,v_2 ,\cdots,v_n )$  with variable ordering $v_1  < v_2  < \cdots < v_n$. Define $g$ such that if $f_{v_k }  = R_x (\alpha _1 )f_{\overline v _k } $ then $g = \hat 1$; otherwise $g = \hat 0$. Assume that $v_{k + 1} ,v_{k + 2} ,\cdots,v_n $ are r-linear variables of $f$ and $v_k$ is a r-nonlinear variable of $f$ with r-deg$(v_k)=m-1$. Additionally, for each value assignment to variables $v_1 , \cdots,v_{k - 1} ,v_{k + 1} , \cdots,v_n$ suppose exactly one of the following $m$ relations holds: $f_{v_k }  = R_x (\alpha _1 )f_{\overline v _k }$, $f_{v_k }  = R_x (\alpha _2 )f_{\overline v _k }$, $\cdots$, $f_{v_k }  = R_x (\alpha _m )f_{\overline v _k } $. We have
\begin{enumerate}[I]
    \item $f$ can be bi-decomposed as $f = g_1 R_x (\gamma )h$ where $g_1  = v_k R_x (\pi )g$, $\gamma  = (\alpha _2  - \alpha _1 )/2$, $h = g_1 R_x ( - \gamma )f$.
    \item $g_1$ is a function of $v_1 ,v_2 ,\cdots,v_k $, i.e., $g_1$ is invariant with respect to $v_{k + 1} ,v_{k + 2} ,\cdots,v_n$.
    \item $v_k$ is a r-linear variable of $g_1$.
    \item $h$ is a function of $v_1 ,v_2 ,\cdots,v_n $ and $v_{k + 1} ,v_{k + 2} ,\cdots,v_n$ are r-linear variables of $h$.
    \item r-deg$(v_k)$ in $h$ is $\leq m-2$.
\end{enumerate}
\end{theorem}

{\bf Proof.} We initially prove that function $g$ is invariant with respect to $v_{k + 1} ,v_{k + 2} ,\cdots,v_n$, i.e., $g_{v_i}  = g_{\overline v _i }$ for $k+1 \leq i \leq n$. Since $v_i$ is r-linear, there exists $\theta_i$ such that for all $v_1 ,\cdots,v_{i - 1} ,v_{i + 1} ,\cdots,v_n $ values, $f_{v_i}  = R_x (\theta _i )f_{{\overline v } _i} $ which results in $f_{v_i v_k }  = R_x (\theta _i )f_{\overline v _i v_k }$ and $f_{v_i \overline v _k }  = R_x (\theta _i )f_{\overline v _i \overline v _k }$. From the definition of $g$ we have:

\begin{itemize}
  \item If $f_{v_i v_k }  = R_x (\alpha _1 )f_{v_i \overline v _k } $, then $g_{v_i }  = \hat 1$, else $g_{v_i }  = \hat 0$.
  \item If $f_{\overline v _i v_k }  = R_x (\alpha _1 )f_{\overline v _i \overline v _k } $, then $g_{\overline v _i }  = \hat 1$, else $g_{\overline v _i }  = \hat 0$.
\end{itemize}

Combining these relations proves $g_{v_i }  = g_{\overline v _i } $:

$$
f_{v_i v_k }  = R_x (\alpha _1 )f_{v_i \overline v _k }  \Leftrightarrow R_x (\theta _i )f_{\overline v _i v_k }  = R_x (\alpha _1  + \theta _i )f_{\overline v _i \overline v _k }  \Leftrightarrow f_{\overline v _i v_k }  = R_x (\alpha _1 )f_{\overline v _i \overline v _k }
$$

Since $g_1  = v_k R_x (\pi )g$, $g_1$ is also invariant with respect to $v_{k + 1} ,v_{k + 2} ,\cdots,v_n$ (part II).  Moreover $g_{1 {v_k }}  = R_x (\pi )g$ and $g_{1 {\bar v_k }}  = g$ which results in $g_{1{v_k }}  = R_x (\pi )g_{1 {\bar v_k }} $, i.e., $v_k$ in $g_1$ is r-linear (part III).

The first sentence of part IV is clear from the definition of $h = g_1 R_x ( - \gamma )f$. As for the second one, note that $v_{k+1}, v_{k+2}, \cdots, v_n$ are r-linear variables of $f$. Additionally, $g$ is invariant with respect to $v_{k + 1} ,v_{k + 2} ,\cdots,v_n$. Putting these facts together proves part IV.

Now we prove r-deg$(v_k)\leq m-2$ in $h = g_1 R_x ( - \gamma )f$. For each value assignment to variables $v_1 ,v_2 ,\cdots,v_{k - 1} ,v_{k + 1} ,\cdots,v_n $ exactly one of the following $m$ relations holds: $f_{v_k }  = R_x (\alpha _1 )f_{\overline v _k } $, $f_{v_k }  = R_x (\alpha _2 )f_{\overline v _k } $, $\cdots$, $f_{v_k }  = R_x (\alpha _m )f_{\overline v _k }$. For each of the above cases, we examine the relation between $h_{v _k}$ and $h_{{\overline v } _k}$:

\begin{itemize}
  \item $f_{v_k }  = R_x (\alpha _1 )f_{\overline v _k }$: By definition $g=\hat 1$ and we have:
\[
\begin{array}{l}
 h_{{\overline v } _k}  = \hat 1 R_x ( - \gamma )f_{{\overline v } _k}  = R_x ( - \gamma )f_{{\overline v } _k}  \Rightarrow f_{{\overline v } _k}  = R_x (\gamma )h_{{\overline v } _k}  \\
 h_{v _k}  = [R_x (\pi )\hat 1]R_x ( - \gamma )f_{v _k}  = f_{v _k}  = R_x (\alpha _1 )f_{{\overline v } _k}  = R_x (\alpha _1  + \gamma )h_{{\overline v } _k}, \gamma  = (\alpha _2  - \alpha _1 )/2 \Rightarrow \\
 h_{v _k}  = R_x (\frac{{\alpha _1  + \alpha _2 }}{2})h_{{\overline v } _k}  \\
 \end{array}
\]
\item $f_{v_k }  = R_x (\alpha _2 )f_{\overline v _k }$: By definition $g=\hat 0$ and we have:
\[
\begin{array}{l}
 h_{{\overline v } _k}  = \hat 0R_x ( - \gamma )f_{{\overline v } _k}  = f_{{\overline v } _k}  \\
 h_{v _k}  = [R_x (\pi )\hat 0]R_x ( - \gamma )f_{v _k}  = R_x ( - \gamma )f_{v _k}  = R_x ( - \gamma  + \alpha _2 )f_{{\overline v } _k}, \gamma  = (\alpha _2  - \alpha _1 )/2 \Rightarrow   \\
 h_{v _k}  = R_x (\frac{{\alpha _1  + \alpha _2 }}{2})h_{{\overline v } _k}  \\
 \end{array}
\]
\item $f_{v_k }  = R_x (\alpha _i )f_{\overline v _k }$, $3\leq i \leq m$: By definition $g=\hat 0$ and $h_{v _k}  = R_x ( - \gamma  + \alpha _i )h_{{\overline v } _k}$.
 \end{itemize}

The first two cases result in the same relation between $h_{v_k}$ and $h_{{\overline v } _k}$ as $h_{v _k} = R_x (\frac{{\alpha _1  + \alpha _2 }}{2}) h_{{\overline v } _k}$. The remaining $m-2$ cases result in at most $m-2$ different relations between $h_{v_k}$ and $h_{{\overline v } _k}$. Therefore, the total number of different relations between $h_{v_k}$ and $h_{{\overline v } _k}$ is $\leq m-1$. Accordingly, r-deg$(v_k)$ in $h$ is $\leq m-2$ (part V).

Finally, from $h = g_1 R_x ( - \gamma )f$ it follows that $g_1 R_x (\gamma )h = g_1 R_x (\gamma )\left[ {g_1 R_x ( - \gamma )f} \right]$. Consider $g_1  = v_k R_x (\pi )g$ and assume $g=\hat 1$ (or $g=\hat 0$) which leads to $g_1 = v_k R_x(\pi) \hat 1=\sim v_k$ (or $g_1 = v_k R_x(\pi) \hat 0=v_k$). Altogether for both $v_k=\hat 1$ and $v_k=\hat 0$, we have $g_1 R_x (\gamma )\left[ {g_1 R_x ( - \gamma )f} \right]=f$. Hence, $f$ can be bi-decomposed as $f = g_1 R_x (\gamma )h$ (part I).
\qed

%\vspace{0.5cm}
Using the proposed bi-decomposition approach, $f$ can be bi-decomposed into $f = g_1 R_x (\gamma )h$ where $g_1$ and $h$ are themselves recursively bi-decomposed until a rotation-based factored form is obtained.

\begin{theorem} \label{theorem:2}
\normalfont
  The proposed bi-decomposition approach always results in a cascade expression for a given function $f$.
\end{theorem}

{\bf Proof.} Following the definitions given in Theorem \ref{theorem:1} for $f = g_1 R_x (\gamma )h$, since $g_1$ is invariant of $v_{k + 1} ,v_{k + 2} ,\cdots,v_n$ and $v_k$ in $g$ is r-linear and r-deg$(v_k)$ in $h$ is $\leq m-2$, the recursion will finally stop at terminal cases where $g_1$ and/or $h$ have directly realizable RbDDs --- all variables are r-linear in the functions and they have rotation-based cascade expressions corresponding to RbDDs with a chain structure.\footnote{As a result of Lemma \ref{lemma:1}, in a function with chain structured RbDD, all variables are r-linear.}\qed

%\vspace{0.5cm}
Algorithm \ref{Alg:qfact} uses the proposed recursive bi-decomposition approach to generate a rotation-based factored form for a given function $f$. All steps in Algorithm \ref{Alg:qfact} can be directly performed on RbDDs. If the RbDD of a function $f$ is a chain structure, we have a cascade expression for $f$ (Step 1). For Step 2 as depicted in Fig. \ref{Fig:q_linear} and according to Lemma \ref{lemma:1}, identifying $v_k$ is equivalent to identifying the lower chain-structure part of the RbDD. As for Step 3, according to Lemma \ref{lemma:2} values $\alpha _1 ,\alpha _2 ,\cdots,\alpha _m $ can be obtained from weights of the $\hat 1$-edges of nodes with decision variables $v_k$. Hence, $\gamma  = (\alpha _2  - \alpha _1 )/2$ is obtained. Let $n_i (1 \leq i \leq m)$ denote nodes with decision variable $v_k$ and $\hat 1$-edges weight $R_x(\alpha_i)$. Starting from the RbDD of $f$, one can perform Algorithm \ref{Alg:g1fact} to construct RbDD of $g_1$. Having the RbDDs for $g_1$ and $f$, the RbDD of $h = g_1 R_x ( - \gamma )f$ can be obtained by using the \texttt{apply} operation. As an example of Algorithm \ref{Alg:g1fact}, see RbDDs of $s$ and $g_1$ in Fig. \ref{Fig:Four-Toffoli} and Fig. \ref{Fig:Four-Toffoli-RbDDs} where $v_k=c$ and $m=2$. {This example is described in detail in Section \ref{sec:ex}.}

The final form after \texttt{apply} is $\mathop {f = g_1 R_x (\gamma _1 )\left[ {g_2 R_x (\gamma _2 )\left[ {g_3 R_x (\gamma _3 ){\rm  }\cdots{\rm   }\left[ {g_k R_x (\gamma _k )\hat 0} \right]{\rm  }} \right]{\rm  }} \right]}\limits^{} $. Note that $g_i$ functions should also be decomposed. The \texttt{factor} algorithm is not optimal. In particular, $f$ can be rewritten as $f = g_{p _1} R_x (\gamma _{p _1} )\left[ {g_{p _2} R_x (\gamma _{p _2} )\left[ {g_{p _3} R_x (\gamma _{p _3} ){\rm  }\cdots{\rm   }\left[ {g_{p _k} R_x (\gamma _{p _k} )\hat 0} \right]{\rm  }} \right]{\rm  }} \right]$ where $(p_1 ,p_2 ,\cdots,p_k )$ is a permutation of $1,2, \cdots,k$. Different permutations of $1,2, \cdots,k$ may result in different number of gates after synthesis. {For example, consider the RbDDs of the output $s$ in a 4-input Toffoli gate, shown in Fig. \ref{Fig:Four-Toffoli}, for two different variable orderings $a>b>c>d$ and $a>b>d>c$. In Fig. \ref{Fig:Four-Toffoli2}, $d$ is r-linear. However, none of the variables in Fig. \ref{Fig:Four-Toffoli3} are r-linear. Accordingly, the proposed approach results in fewer gates for $a>b>c>d$. The former case is further discussed in Section \ref{sec:ex}. {Indeed, working with $a>b>d>c$ leads to $s=g_1R_x(-\pi/2)h$ where $g_1=abd \oplus c$ is a 4-input Toffoli gate targeted on the last qubit $c$ for $a>b>d>c$.}}

\algsetup{indent=2em}
\begin{algorithm}[tb]
\caption{\texttt{$g_1$-factor(RbDD$_f$)}\label{Alg:g1fact}}

\begin{algorithmic}[1]
\medskip
\STATE Change all of the weights to $R_x(0)=I$.
\STATE Create a RbDD node $v_k$ with $w_1=R_x(\pi)$ and $w_0=I$ to the terminal node (i.e., $\hat 0$).
\STATE Redirect all edges toward $n_1$ to node $v_k$ and make the weight of all such edges $R_x(\pi)$.
\STATE Redirect all edges toward $n_2, n_3, \cdots, n_m$ to node $v_k$ and make the weight of all such edges  $R_x(0)$.
\STATE Discard nodes $n_2, n_3, \cdots, n_m$.
\STATE Merge isomorphic sub-graphs, eliminate nodes with the same $\hat 0$-child and $\hat 1$-child exactly if the weight of the $\hat 1$-edge is $R_x(0)=I$. Update weights of the RbDD to make the RbDD of $g_1$ canonical.
\end{algorithmic}
\end{algorithm}

\section{Working with Arbitrary Outputs} \label{sec:extension}
\begin{figure}
    \centering
    \fbox{
    \subfigure[\label{Fig:GCirc}]{
    \includegraphics[height=1.4cm]{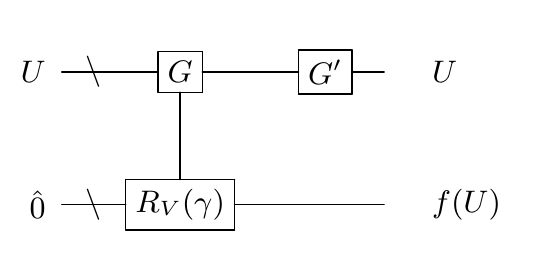}
    }
    \subfigure[\label{Fig:GMux}]{
    \includegraphics[height=1.7cm]{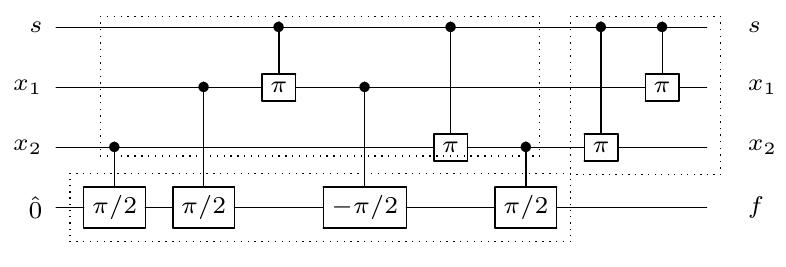}
    }
    \subfigure[\label{Fig:GCirc2}]{
    \includegraphics[height=1.4cm]{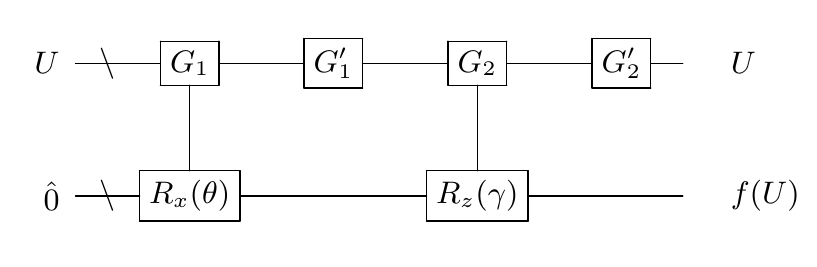}
    }
    }
    \vspace{2mm}
    \caption{(a) Quantum circuit performing $f(U) = R_x ({\gamma (U)} ) \hat 0$. (b) Quantum 3-input multiplexer. Dashed boxes represent $G$, $G'$, and $R_x(\gamma)$ in (a). Only rotation angles are reported for $R_x(\theta)$ gates. (c) Quantum circuit performing $f(U) = R_z (\gamma (U))R_x (\theta (U))\hat 0$.}
\end{figure}

\begin{figure}
    \centering
    \fbox{
    \subfigure[\label{Fig:Four-Toffoli1}]{
    \includegraphics[height=2cm]{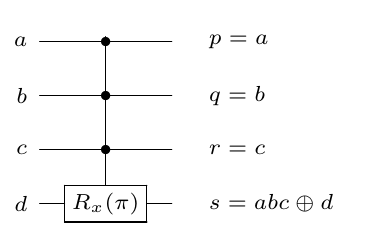}
    }
    \subfigure[\label{Fig:Four-Toffoli2}]{
    \includegraphics[height=4.5cm]{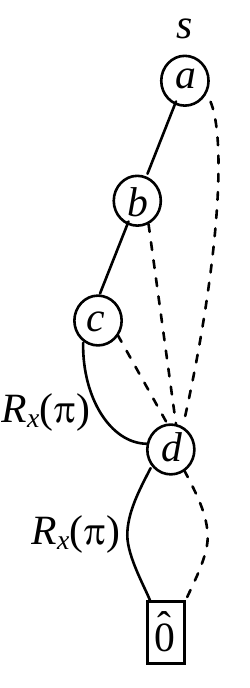}
    }
    \subfigure[\label{Fig:Four-Toffoli3}]{
    \includegraphics[height=4.5cm]{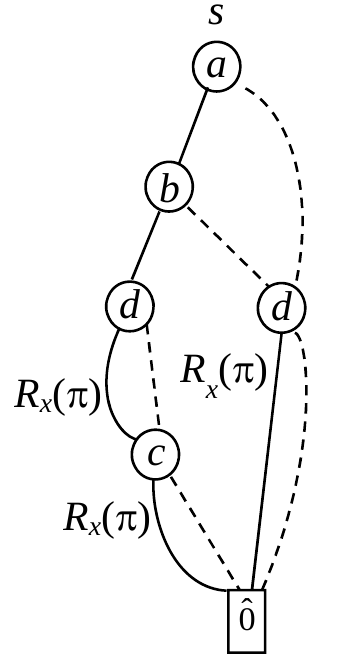}
    }
    }
    \vspace{2mm}
    \caption{\label{Fig:Four-Toffoli} 4-input Toffoli gate (a), the RbDD for the 4th output $s$ in two cases: if $a > b > c> d$ (b){, and if $a>b>d>c$ (c)}.}
\end{figure}

For the input vector $U$, a function $f$ with binary inputs and outputs can be written as $f(U) = \hat g_1 (U)R_x (\gamma _1 )$ $\left[ {\hat g_2 (U)R_x (\gamma _2 )\left[ {{\rm  } \cdots {\rm   }\left[ {\hat g_k (U)R_x (\gamma _k )\hat 0} \right]{\rm  }} \right]{\rm  }} \right]$. Since functions $\hat g_i(U)$ only take values $\hat 0$ and $\hat 1$, $f(U)$ can also be represented as $f(U) = R_x ( {g_1 (U)\gamma _1  + g_2 (U)\gamma _2  + \cdots + g_k (U)\gamma _k } )\hat 0$ where $g_i(U)$ values are either zero (0) or one (1).\footnote{To prove, assign arbitrary values $\hat 0$ and $\hat 1$ to $\hat g_i$ terms and consider the resulting rotations by different $\gamma_i$ values.} Define $\gamma (U) = g_1 (U)\gamma _1  + g_2 (U)\gamma _2  + \cdots + g_k (U)\gamma _k $ which leads to $f(U) = R_x ({\gamma (U)} )\hat 0$. Accordingly, the structure of the synthesized circuit can be represented as Fig. \ref{Fig:GCirc}. In this figure, $G$ is a circuit that constructs $\gamma(U)$ and $G'$ is the inverse of $G$. Note that $G'$ should be used only if one wants to keep input lines unchanged. To clarify the roles of $G$ and $G’$, see the 3-input multiplexer circuit $f = sx_1  + \bar sx_2 $ synthesized by the \texttt{factor} algorithm in Fig. \ref{Fig:GMux}. If instead of $\hat 0$, another quantum value $q$ is used in this circuit as the initial value for the input, the resulting circuit implements $f(U) = R_x \left[ {\gamma (U)} \right]q$. The constant ancilla register in Fig. \ref{Fig:GCirc} may not be necessary in some case. For example, the controlled rotation $R_x (\pi )$ with control qubit $a$ and target $\hat 0$ generates $a$ as the second output and the use of the controlled rotation $R_x (\pi )$ in this case is unnecessary (i.e., $aR_x (\pi )\hat 0=a$). Section \ref{sec:ex} shows several examples.

Now consider a given function that for given basis input vectors generates a general value $f(U) = \left[ {\begin{array}{*{20}c} {f_0 (U)} & {f_1 (U)}  \\ \end{array}} \right]^T $. Since $|f_0 (U)|^2  + |f_1 (U)|^2  = 1$, we may rewrite $f(U)$ as:

$$
f(U) = e^{i\delta (U)} \left[ {\begin{array}{*{20}c} {e^{ - i\gamma (U)/2} \cos \frac{{\theta (U)}}{2}} & { - ie^{i\gamma (U)/2} \sin \frac{{\theta (U)}}{2}}  \\ \end{array}} \right]^T
$$

Hence, $f(U)$ can be expressed as $f(U) = e^{i\delta (U)} R_z (\gamma (U))R_x (\theta (U))\hat 0$ where $R_z$ is the rotation operator around the $z$ axis. We can ignore the global phase $e^{i\delta (U)}$ since it has no observable effects \cite{NeilsenChuang}. Therefore, one can effectively write $f(U) = R_z (\gamma (U))R_x (\theta (U))\hat 0$. Note that $R_z (\gamma )R_x (\theta )\hat 0$ results from $\theta$ rotation of $\hat 0$ around the $x$ axis followed by $\gamma$ rotation around the $z$ axis in the Bloch sphere. The quantum circuit for $f(U) = R_z (\gamma (U))R_x (\theta (U))\hat 0$ can be synthesized as:
\begin{itemize}
\item Synthesize $g(U) = R_x (\theta (U))\hat 0$ by using the \texttt{factor} algorithm.
\item Synthesize $h(U)=R_z(\gamma(U))\hat 0$ by using the \texttt{factor} algorithm.
\item Cascade the resulting circuits as depicted in Fig. \ref{Fig:GCirc2}. In this figure, $G_1$ and $G_2$ are for $g(U)$ and $h(U)$, respectively. Accordingly, $G_1'$ and $G_2'$ are the inverse circuits of $G_1$ and $G_2$.
\end{itemize}

\section{Results} \label{sec:ex}
%\begin{example} \label{ex:Toffoli}
\normalfont
{\bf Multiple-control Toffoli gate.} Consider a \textbf{4-input Toffoli} gate in Fig. \ref{Fig:Four-Toffoli1} and the RbDD of the target output in Fig. \ref{Fig:Four-Toffoli2} with variable ordering $a<b<c<d$. Comparing the RbDD of $s$ with the general RbDD structure in Fig. \ref{Fig:q_linear} reveals that variable $c$ corresponds to $v_k$. Additionally, r-deg$(c)$=1, $\alpha_1=0$ and $\alpha_2=\pi$ which result in $\gamma=\pi/2$ (Theorem \ref{theorem:1}).\footnote{One may set $\alpha_1=\pi$ and $\alpha_2=0$. This combination generates a different circuit with the same functionality.}

\begin{figure}
    \centering
    \fbox{
    \includegraphics[height=6cm]{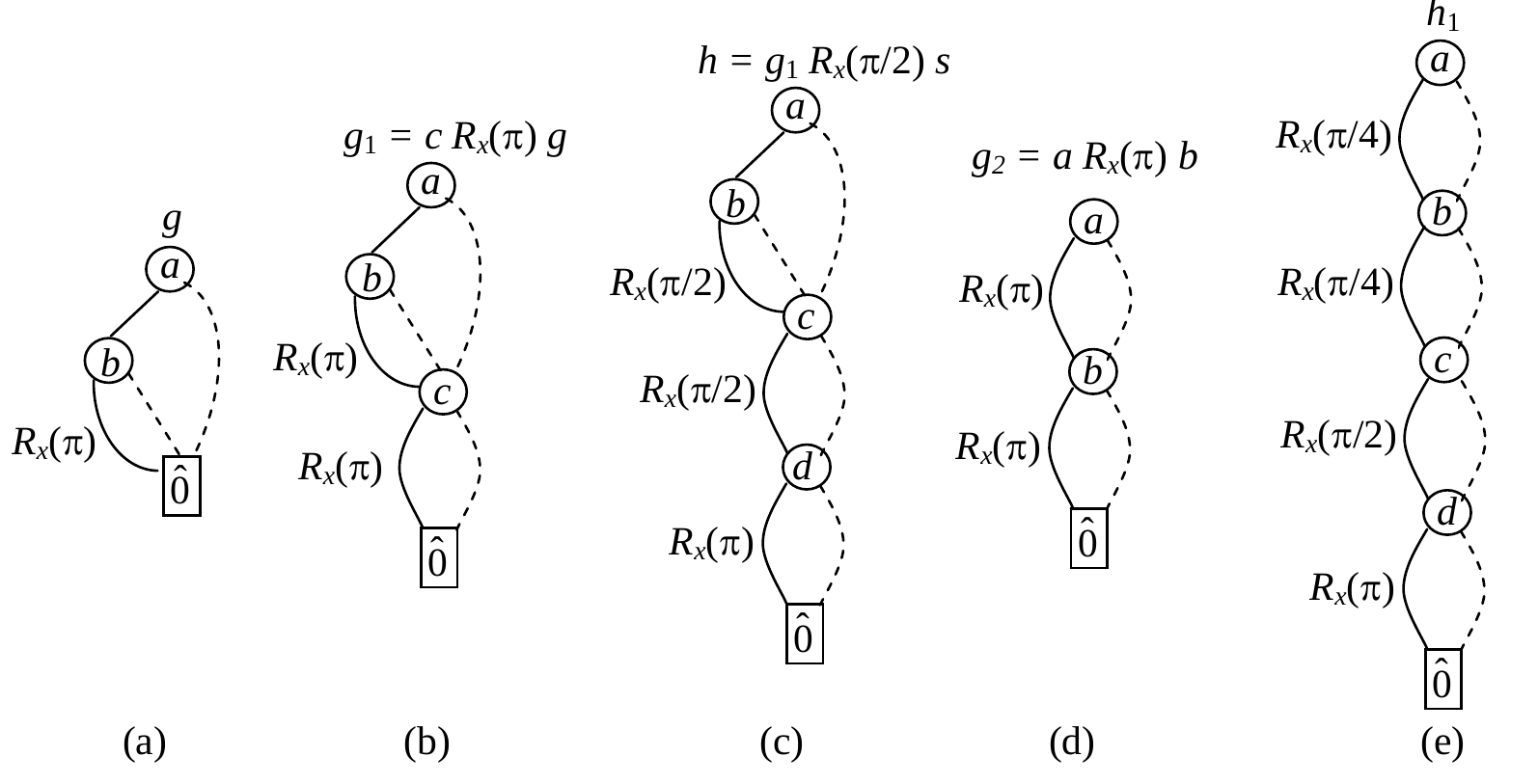}
    }
    \vspace{2mm}
    \caption{\label{Fig:Four-Toffoli-RbDDs} RbDDs required to synthesize the 4-input Toffoli gate in Fig. \ref{Fig:Four-Toffoli}.}
\end{figure}

Function $s$ can be bi-decomposed as $s = g_1 R_x ( - \pi /2)h$ where $g_1  = cR_x (\pi )g$. RbDDs for $g$ and $g_1$ are shown in Fig. \ref{Fig:Four-Toffoli-RbDDs}(a) and Fig. \ref{Fig:Four-Toffoli-RbDDs}(b), respectively.\footnote{RbDD of $g_1$ can be obtained by using Algorithm \ref{Alg:g1fact} and RbDD of $s$ --- no need to construct RbDD of $g$. However, an interested reader can verify that an indirect approach to construct RbDD of $g$ (and hence the function of $g$) is to replace $v_k$ by $\hat 0$ in RbDD of $g_1$ which is constructed from applying Algorithm \ref{Alg:g1fact}.} Note that $g_1$ is a 3-input Toffoli gate (see RbDD of $r$ in Fig. \ref{Fig:qapply}), which can be synthesized as in Fig. \ref{Fig:CTGates}(c). As for function $h$, it can be written as $h = g_1 R_x (\pi /2)s$. The RbDD for $h$ (by the \texttt{apply} operator) is shown in Fig. \ref{Fig:Four-Toffoli-RbDDs}(c). Subsequently, $h$ can be bi-decomposed as $h = g_2 R_x ( - \pi /4)h_1$ where $g_2  = aR_x (\pi )b$ (by algorithm \ref{Alg:g1fact}) and $h_1  = g_2 R_x (\pi /4)h$ (by the \texttt{apply} operator). The resulting RbDDs for $g_2$ and $h_1$ are shown in Fig. \ref{Fig:Four-Toffoli-RbDDs}(d) and Fig. \ref{Fig:Four-Toffoli-RbDDs}(e). Finally, the factored form for $s$ is $s = g_1 R_x ( - \pi /2)[g_2 R_x ( - \pi /4)h_1 ]$.

Due to the chain structure of $g_2$ and $h_1$, they may be directly realized by using controlled-rotation operators. Note that when realizing $g_1$, we also implement $g_2$. The final circuit is shown in Fig. \ref{Fig:Four-Toffoli-Circuit}. The first subcircuit generates output $s$ whereas the remaining gates generate outputs $p$, $q$ and $r$.

%\vspace{0.5cm}
As a direct extension of the above approach, consider a \textbf{multiple-control Toffoli gate }on $n+1$ qubits with controls $i_1, i_2, ..., i_n$ and target $j$. Toffoli output can be written as $j = i_1 i_2 \cdots i_n \oplus j$. Assume $i_1 < i_2 < \cdots < i_n < j$. It can be verified that $v_k$ (in Algorithm \ref{Alg:qfact}) is $i_n$ and we have r-deg$(i_n)=1$ with $\alpha_1=0$, $\alpha_2=\pi$, and $\gamma=\pi/2$ (in Theorem \ref{theorem:1}). Therefore, one can write $j=g_1 R_x(-\pi/2) h$. It results in $g_1 = i_1 i_2 \cdots i_{n-1} \oplus i_n$ and $h= [i_1 i_2 \cdots i_{n-1}] R_x(\pi/2) [i_n R_x(\pi/2) j]$. Now, $g_1$ is an $n$-qubit Toffoli gate and can be decomposed independently following the same approach. To decompose $h$, one can verify that $v_k=i_{n-1}$ in Algorithm \ref{Alg:qfact} with r-deg$(i_{n-1})=1$, $\alpha_1=0$, $\alpha_2=\pi/2$, and $\gamma=\pi/4$. Accordingly, we can write $h = g_2 R_x(-\pi/4) h_1$. Applying Algorithm \ref{Alg:g1fact} reveals that $g_2$ is an $(n-1)$-qubit Toffoli gate with $i_{n-1}$ as the target and $i_1, i_2, \cdots, i_{n-2}$ as controls. By using the \texttt{apply} operator, $h_1=g_2 R_x(\pi/4) h$ which leaves $v_k=i_{n-2}$. Altogether, we can write:

\begin{align*}
C^nR_x(\pi) &= C^{n-1}R_x(\pi) R_x(-\pi/2)  \nonumber \\
     &\qquad \,\,\,\,\,\, [C^{n-2}R_x(\pi) R_x(-\pi/4)  \nonumber \\
      &\qquad \,\,\,\,\,\,\,\,\,\,\,\,  [C^{n-3}R_x(\pi) R_x(-\pi/8) \nonumber \\
       &\qquad \,\,\,\,\,\,\,\,\,\,\,\,\,\,\,\,\,\,     [\cdots  \nonumber \\
       &\qquad \,\,\,\,\,\,\,\,\,\,\,\,\,\,\,\,\,\,\,\,\,\,\,\,     [C^{n-(n-1)}R_x(\pi) R_x(-\pi/2^{n-1}) \nonumber \\
        &\qquad         [i_1 R_x(\pi/2^{n-1})(i_2 R_x(\pi/2^{n-1}) (i_3 R_x(\pi/2^{n-2})(\cdots (i_n R_x(\pi/2) j ) \cdots )))             ]\nonumber \\
        &\qquad             ] \cdots ]]]\nonumber \\
\end{align*}

{To construct the circuit, for $[i_1 R_x(\pi/2^{n-1})(i_2 R_x(\pi/2^{n-1}) (\cdots (i_n R_x(\pi/2) j ) \cdots ))]$ one needs to add $n$ controlled-rotation gates with controls on $i_1$, $i_2$, $\cdots$, $i_n$ and targets on $j$. This subcircuit should be followed by constructing a $g_1=C^{n-1}R_x(\pi)$ gate which automatically constructs all $g_2=C^{n-2}R_x(\pi)$, $g_3=C^{n-3}R_x(\pi)$, $\cdots$, $g_n=CR_x(\pi)$ gates too. Next, one needs to use $n-1$ controlled-rotation gates with controls on $g_2$, $g_3$, $\cdots$, $g_n$ and targets on $j$. Altogether, we need $\rm{COST}_{1,C^{n}\rm{NOT}}=2n-1+\rm{COST}_{1,C^{n-1}\rm{NOT}}$ controlled-rotation gates to implement a {C$^{n}$NOT} gate. To restore $i_1$, $i_2$, $\cdots$, $i_n$ qubits to their original values, additional cost should be applied which is $\rm{COST}_{2,C^{n}\rm{NOT}}=\rm{COST}_{1,C^{n-1}\rm{NOT}}$, i.e., all gates excluding gates with targets on $j$. Terminal conditions are $\rm{COST}_{1,C^2\rm{NOT}}=4$ and $\rm{COST}_{2,C^2\rm{NOT}}=1$ (see Fig. \ref{Fig:CTGates}(c)). Total implementation cost is $\rm{COST}_{C^{n}\rm{NOT}}=\rm{COST}_{1,C^{n}\rm{NOT}}+\rm{COST}_{2,C^{n}\rm{NOT}}$ which is \emph{polynomial}, i.e., $2n^2-2n+1$. Fig. \ref{Fig:Five-Toffoli-Circuit} illustrates this construction for a 5-input Toffoli gate. No ancilla is required in the proposed construction.} {Current constructions for a $C^{n}\rm{NOT}$ gate use an \emph{exponential} number of 2-qubit gates $2^{n+1}-3$ \cite[Lemma 7.1]{Barenco95} or $48n^2 + O(n)$ arbitrary 2-qubit operations \cite[Lemma 7.6]{Barenco95}, if no ancilla is available.} \qed
%\end{example}

\begin{figure}[tb]
    \scriptsize
    \centering
    \input{figures/Four-Toffoli}
    \vspace{2mm}
    \caption{\label{Fig:Four-Toffoli-Circuit} Automatic synthesis of a 4-input Toffoli gate obtained by the \texttt{factor} algorithm. Only rotation angles are reported for $R_x(\theta)$ gates. The first subcircuit generates output $s$ in Fig. \ref{Fig:Four-Toffoli} whereas the remaining gates generate outputs $p$, $q$ and $r$. }
\end{figure}

\begin{figure}[tb]
    \scriptsize
    \centering
    \scalebox{0.75}{
    \input{figures/Five-Toffoli}
    }
    \vspace{2mm}
    \caption{\label{Fig:Five-Toffoli-Circuit} {Synthesized circuit for a 5-input Toffoli gate. Only rotation angles are reported for $R_x(\theta)$ gates. The dashed subcircuit generates output for a 4-input Toffoli gate (see Fig. \ref{Fig:Four-Toffoli-Circuit}) and $U^{-1}$ is the reverse of this subcircuit.}}
\end{figure}

\begin{figure}[tb]
    \centering
    \fbox{
    \subfigure[\label{Fig:FAQDDs}]{
    \includegraphics[height=4.8cm]{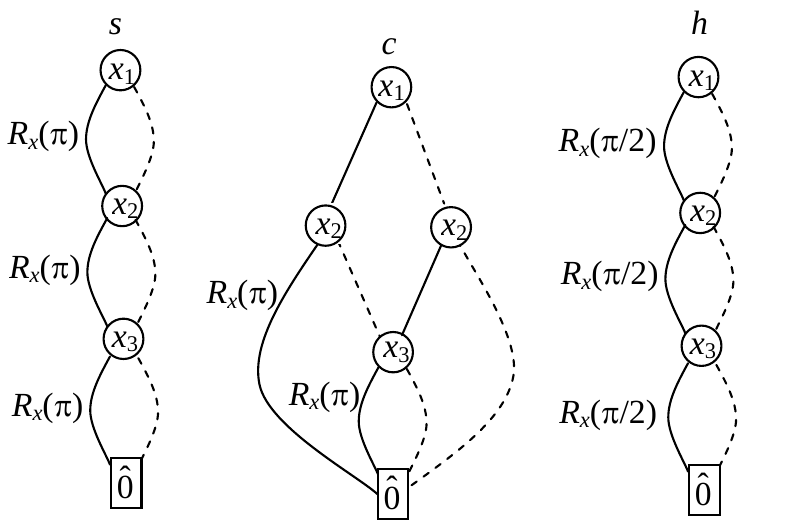}
    }
    \subfigure[\label{Fig:FACirc}]{
    \includegraphics[height=2.3cm]{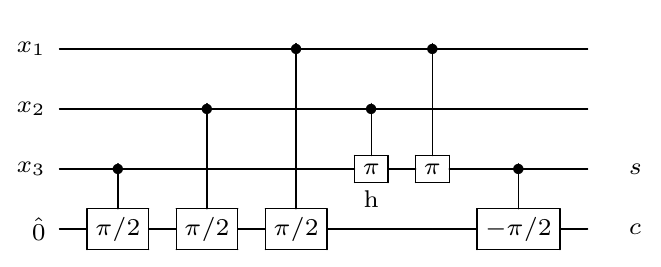}
    }
    }
    \vspace{2mm}
    \caption{(a) RbDDs for a 3-qubit full adder and (b) the synthesized circuit. Only rotation angles are reported for $R_x(\theta)$ gates. }
\end{figure}

\vspace{0.3cm}
%\begin{example}  \label{ex:FA}
\normalfont
{\bf Quantum adder.} Consider a \textbf{full adder} with inputs $x_1$, $x_2$, and $x_3$ ($x_1 < x_2 < x_3$) and outputs $s = x_1  \oplus x_2  \oplus x_3$ and $c = x_1 x_2  + x_1 x_3  + x_2 x_3$. The RbDDs of $s$ and $c$ are shown in Fig. \ref{Fig:FAQDDs}. The RbDD of $s$ has a chain structure that corresponds to a cascade expression and can be directly realized. On the other hand, the RbDD of $c$ should be recursively decomposed by using Algorithm \ref{Alg:qfact}. Using this algorithm, $c$ is bi-decomposed as $c = g_1 R_x ( - \pi /2)h$.

To construct RbDD of $g_1$ note that $v_k=x_3$. Applying Algorithm \ref{Alg:g1fact} leads to four internal nodes as follows. Node $n_1$ with the decision variable $x_1$, $w_1=w_0=I$, and $\hat 1$-child node $n_2$, and $\hat 0$-child node $n_3$. Node $n_2$ with the decision variable $x_2$, node weight $R_x(\pi)$, $w_1=R_x(\pi)$, $w_0=I$, and node $n_4$ as both $\hat 1$-child and $\hat 0$-child. Node $n_3$ with the decision variable $x_2$, $w_1=R_x(\pi)$, $w_0=I$, and node $n_4$ as both $\hat 1$-child and $\hat 0$-child. Node $n_4$ with decision variable $x_3$ connected to the terminal node $\hat 0$ with $w_1=R_x(\pi)$, and $w_0=I$. A careful consideration reveals that this RbDD can be converted to the one constructed for $s$ in Fig. \ref{Fig:FAQDDs}. Therefore, $g_1$ has a cascade expression and a realizable rotation-based implementation. Finally, the RbDD for $h = g_1 R_x (\pi /2)c$ is shown in Fig. \ref{Fig:FAQDDs}. As can be seen, the RbDD of $h$ has a chain structure too. The resulting quantum circuit is depicted in Fig. \ref{Fig:FACirc}.

Now consider a \textbf{2-qubit quantum adder} with inputs $a_1$, $a_0$, $b_1$, $b_0$ for $a_0 < b_0 < a_1 < b_1$ and outputs $c$, $s_1$, and $s_0$ for $s_0 < s_1 < c$. It can be verified that $s_0=b_0 \oplus a_0, s_1= a_0b_0 \oplus a_1 \oplus b_1, c = a_0b_0a_1 \oplus a_0 b_0 b_1 \oplus a_1 b_1$. Applying the above approach leads to the following equations:

\begin{equation*}
\begin{array}{l}
s_0 = a_0 R_x(\pi) b_0 \\
\left\{ \begin{array}{l}
s_1 = g_1 R_x(-\pi/2) h_1 \\
g_1 = s_0 \\
h_1 = a_0 R_x(\pi/2) (b_0 R_x(\pi/2) (a_1 R_x(\pi) b_1)) \\
 \end{array} \right. \\
\left\{ \begin{array}{l}
c = g_2 R_x(-\pi/2) h_2 \\
g_2 = s_1 \\
h_2 = g_3 R_x(-\pi/4) h_3 \\
g_3 = s_0 \\
h_3 = a_0 R_x(\pi/4) (b_0 R_x(\pi/4) (a_1 R_x(\pi/2) (b_1 R_x(\pi/2) \hat 0))) \\
 \end{array} \right.
 \end{array}
\end{equation*}

Therefore, $s_0, s_1$, and $c$ can be implemented by one, four, and six 2-qubit gates (11 in total), respectively. The circuit uses one ancilla for $c$; $a_0, a_1$ remain unchanged and $s_1$ and $s_0$ are constructed on $b_1$ and $b_0$, respectively.

To generalize, consider an \textbf{$n$-qubit quantum ripple adder} with inputs $a_i$ and $b_i$ and outputs $s_i$ and $c$ for $0 \leq i \leq n-1$ and $a_0 < b_0 < a_1 < b_1 < \cdots < a_{n-1} < b_{n-1}$ and $s_0 < s_1 < \cdots <s_{n-1}< c$. We have:

\begin{equation*}
\begin{array}{l}
s_0 = a_0 R_x(\pi) b_0 \\
s_1 = s_0 R_x(-\pi/2) (a_0 R_x(\pi/2) (b_0 R_x(\pi/2) (a_1 R_x(\pi) (b_1 R_x(\pi) \hat 0)))) \\
\cdots\\
\end{array}
\end{equation*}
\begin{align*}
s_{n-1} &= s_{n-2} R_x(-\pi/2) (s_{n-3} R_x(-\pi/2^2) (\cdots (s_0 R_x(-\pi/2^{n-1}) \nonumber \\
 &\qquad \,\,\,\,\,\, (a_0 R_x(\pi/2^{n-1}) (b_0 R_x(\pi/2^{n-1}) \nonumber \\
 &\qquad \,\,\,\,\,\, (a_1 R_x(\pi/2^{n-2}) (b_1 R_x(\pi/2^{n-2}) \nonumber \\
 &\qquad \,\,\,\,\,\, \cdots  \nonumber \\
 &\qquad \,\,\,\,\,\, (a_{n-2} R_x(\pi/2) (b_{n-2} R_x(\pi/2)) \nonumber \\
 &\qquad \,\,\,\,\,\, (a_{n-1} R_x(\pi) (b_{n-1} R_x(\pi) \hat 0)))\cdots) \nonumber \\
c &= s_{n-1} R_x(-\pi/2) (s_{n-2} R_x(-\pi/2^2) (\cdots (s_0 R_x(-\pi/2^n) \nonumber \\
 &\qquad \,\,\,\,\,\, (a_0 R_x(\pi/2^{n}) (b_0 R_x(\pi/2^{n}) \nonumber \\
 &\qquad \,\,\,\,\,\, (a_1 R_x(\pi/2^{n-1}) (b_1 R_x(\pi/2^{n-1}) \nonumber \\
 &\qquad \,\,\,\,\,\, \cdots  \nonumber \\
 &\qquad \,\,\,\,\,\, (a_{n-2} R_x(\pi/4) (b_{n-2} R_x(\pi/4)) \nonumber \\
 &\qquad \,\,\,\,\,\, (a_{n-1} R_x(\pi/2) (b_{n-1} R_x(\pi/2) \hat 0)))\cdots) \nonumber \\
\end{align*}

\begin{figure}[tb]
    \scriptsize
    \centering
    \scalebox{1}{
    \input{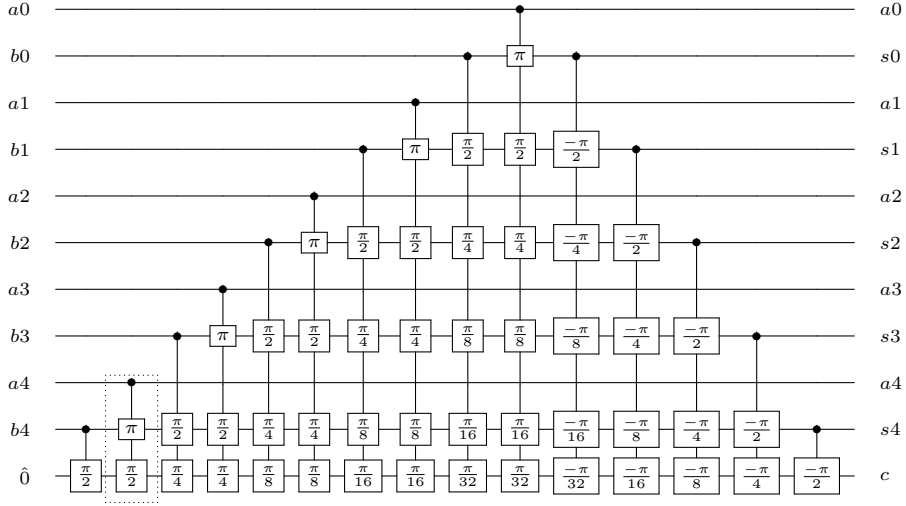}
    }
    \vspace{2mm}
    \caption{\label{Fig:Adder5} {5-bit ripple-carry adder synthesized by the proposed approach with 50 2-qubit rotation gates. Different gates with the same control line are represented as one gate with several targets. For example, the gate in the dashed box includes two rotation gates conditioned on $a_4$ with targets on $c$ (ancilla) and $b_4$.} }
\end{figure}

\begin{figure}[tb]
    \scriptsize
    \centering
    \scalebox{.8}{
    \input{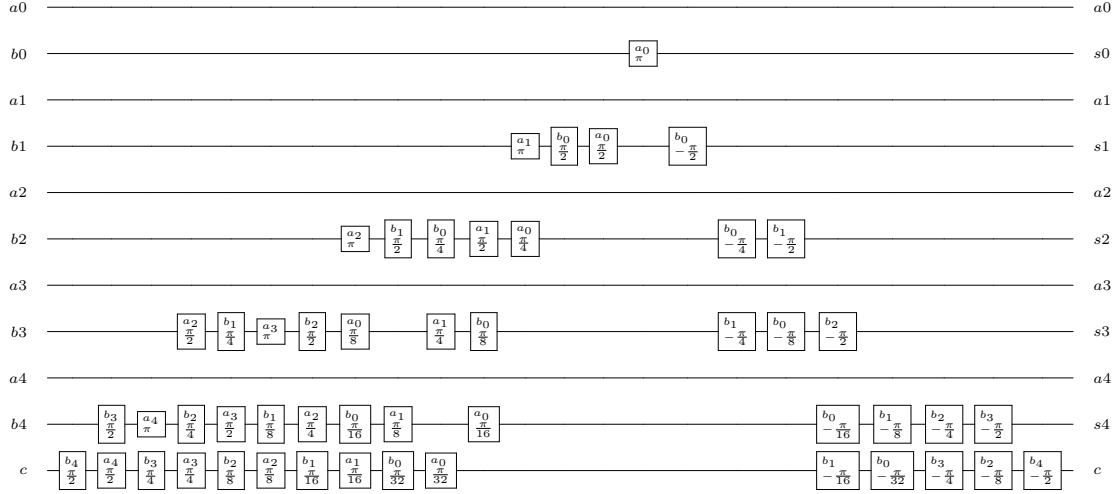}
    }
    \vspace{2mm}
    \caption{\label{Fig:Adder5p} {Circuit in Fig. \ref{Fig:Adder5} with depth 23. The notation \SquareBox{$^{a}_{\theta}$} represents controlled-rotation gates where $a$ is the control line and $\theta$ is the rotation angle.}}
\end{figure}

{To count the number of 2-qubit gates, note that there are $2n$ gates on $c$, $2n-1$ gates on $b_{n-1}$, $2n-3$ gates on $b_{n-2}$, $\cdots$, 3 gates on $b_1$ and 1 gate on $b_0$ in the proposed construction. This subcircuit should be followed by a 2-qubit gate conditioned on $b_0$ with target on $b_1$, 2 gates conditioned on $b_0$ and $b_1$ with targets on $b_2$, 3 gates conditioned on $b_0$, $b_1$, $b_2$ with targets on $b_4$, etc. Altogether, an $n$-qubit quantum ripple adder can be implemented with $1/2(3n^2+5n)$ controlled-rotation gates. Fig. \ref{Fig:Adder5} illustrates the proposed construction for a 5-qubit carry-ripple adder. This circuit is restructured in Fig. \ref{Fig:Adder5p} with parallel gates. Compared to the construction in \cite[Figure 7]{Cuccaro2004} with depth 28, our circuit uses a wider varieties of rotation angles to reduce the depth to 23. Circuit depth for $n=2,\cdots,15$ is 9(10), 12(16), 19(22), 23(28), 27(34), 31(40), 39(46), 43(52), 48(58), 51(64), 57(70), 61(76), 66(82), 70(88) where $a(b)$ denotes $a$ 2-qubit gates in the proposed construction and $b$ 2-qubit gates in \cite{Cuccaro2004}.}\footnote{{The construction in \cite{Cuccaro2004} generates a circuit with controlled-rotation gates with phase $\pi/2$ and total depth $6n-2$ for an $n$-qubit carry-ripple adder. We guess our circuit depth is $5n+O(1)$. The trend line for the number of 2-qubit gates in the proposed construction for $n=2,\cdots,15$ is $4.7868n - 0.9736$}.}
 \qed

%\end{example}

\vspace{0.3cm}
%\begin{example}  \label{ex:mux}
\normalfont
{\bf Quantum multiplexer.} Consider a \textbf{3-input multiplexer} $f = sx_1  + \bar sx_2 $ with $s<x_1<x_2$. Following Theorem \ref{theorem:1} leads to $\alpha_1=0, \alpha_2=\pi, \gamma=\pi/2$ and $f=g_1 R_x(\pi/2) h$ for $v_k=x_2$ with r-deg$(x_2)=1$. To construct $g_1$ note that we use $\alpha_1=0$. It results in a chain structure for $g_1$ with the factored form $g_1=s R_x(\pi) x_2$.

To construct the RbDD of $h=g_1 R_x(-\pi/2)f$ using the \texttt{apply} operator, note that both $g_1$ and $f$ use $s$. Accordingly, one needs to apply Rule 3 which results in $w_1 = [R_x(\pi)x_2]R_x(-\pi/2) x_1$ and $w_0=x_2 R_x(-\pi/2) x_2=x_2 R_x(\pi/2) \hat 0$. Continuing this path results in $h=g_2 R_x(-\pi/2) h_1$, $g_2=s R_x(\pi)x_1$, and $h_1=x_1 R_x(\pi/2) [x_2 R_x(\pi/2) \hat 0]$.

\begin{figure}[tb]
    \scriptsize
    \centering
    \scalebox{.9}{
    \input{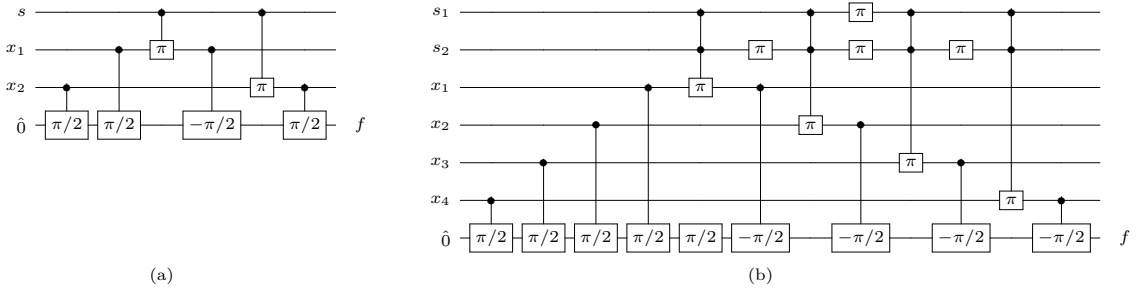}
    }
    \vspace{2mm}
    \caption{\label{Fig:Mux} (a) Quantum 2-to-1 multiplexer, (b) quantum 4-to-1 multiplexer. Circuits are directly obtained by the \texttt{factor} algorithm. Only rotation angles are reported for $R_x(\theta)$ gates. }
\end{figure}

Altogether, $f= [s R_x(\pi) x_2] R_x(-\pi/2) [[s R_x(\pi)x_1] R_x(-\pi/2) [x_1 R_x(\pi/2) [x_2 R_x(\pi/2) \hat 0]]]$. The resulting quantum circuit is shown in Fig. \ref{Fig:Mux}(a). Note that one $\hat 0$-initialized qubit is added to setup $[x_2 R_x(\pi/2) \hat 0]$.

The \textbf{6-input multiplexer} $f = s_1 s_2 x_1  + s_1 \bar s_2 x_2  + \bar s_1 s_2 x_3  + \bar s_1 \bar s_2 x_4$ can be constructed by using three 3-input multiplexers, which would require 3 extra ancillae. However, if the \texttt{factor} algorithm is directly applied, the resulting circuit only uses one ancilla as illustrated in Fig. \ref{Fig:Mux}(b). For an \textbf{$n$-qubit quantum multiplexer} with $\lceil \log n \rceil$ selects and $n$ inputs, the proposed approach leads to $2n$ 2-qubit gates and $n$ C$^{\lceil \log n \rceil}R_x(\pi)$ gates with one ancilla. {As the cost of an $n$-qubit multiple-control Toffoli gate is $2n^2-2n+1$, the proposed approach leads to $O(n\log^2 n)$ gates, i.e., $2n + n(2\lceil \log n \rceil^2-2\lceil \log n \rceil+1)$. We found no explicit construction for an $n$-qubit quantum multiplexer in the literature. However, one can use $n$ $C^{\lceil \log n \rceil+1}$CNOT gates in a circuit with one zero-initialized ancilla to implement an $n$-qubit multiplexer. Considering linear-size cost for each gate \cite{Barenco95} leads to $O(n^2)$ cost.}\footnote{{ For example, a 4-to-1 multiplexer can be implemented as $T(s_0',s_1',x_0,f)$, $T(s_0',s_1,x_1,f)$, $T(s_0,s_1',x_2,f)$, $T(s_0,s_1',x_3,f)$ for a circuit with selects $s_0,s_1$, inputs $x_0, x_1, x_2, x_3$ and output $f$, where $f$ is a zero-initialized ancilla. For each T (Toffoli) gate, the first three lines act as the control lines and the last line acts as the target. In addition, e.g., $T(s_0',s_1',x_0,f)$ applies $x_0$ on $f$ when $s_0=0, s_1=0$. This can be implemented by N($s_0$), N($s_1$), $T(s_0,s_1,x_0,f)$, N$(s_1)$, $N(s_0)$ where N denotes the NOT gate.}}
\qed
%\end{example}

\vspace{0.3cm}
%\begin{example}  \label{ex:qft}
\normalfont
{\bf Quantum Fourier Transform.} The \textbf{quantum Fourier transform (QFT)} ‎is used in many quantum algorithms. QFT has an efficient quantum circuit implementation based on $R_z$ gates \cite{NeilsenChuang}. The result of applying QFT on inputs $j_1 ,j_2 , \ldots ,j_n $ is $\left| 0 \right\rangle  + e^{2\pi i0.j_n } \left| 1 \right\rangle $, $\left| 0 \right\rangle  + e^{2\pi i0.j_2  \cdots j_n } \left| 1 \right\rangle $, $\cdots$, $\left| 0 \right\rangle  + e^{2\pi i0.j_1 j_2  \cdots j_n } \left| 1 \right\rangle $ where the common notation $0.j_1 j_2  \ldots j_n  = \frac{{j_1 }}{2} + \frac{{j_2 }}{{2^2 }} +  \cdots  + \frac{{j_n }}{{2^n }}$ is used.

Following the discussion in Section \ref{sec:extension}, in the first step the output $f_n (j_1 ,j_2 , \ldots ,j_n ) = \left| 0 \right\rangle  + e^{2\pi i0.j_1 j_2  \cdots j_n } \left| 1 \right\rangle $ is described by $f_n (J) = e^{i\delta (J)} R_z (\gamma (J))R_x (\theta (J))\left| 0 \right\rangle $  where $J = j_1 2^{n-1} + j_2 2^{n-2} + \cdots + j_n$. For this function $\theta (J) = (\pi /2)j_1 $, and $g(J) = R_x (\theta (J))\hat 0 = R_x (\pi /2) \hat 0$.

\begin{figure}[tb]
    \centering
    {
    \subfigure[\label{Fig:QFT1}]{
    \includegraphics[height=5.5cm]{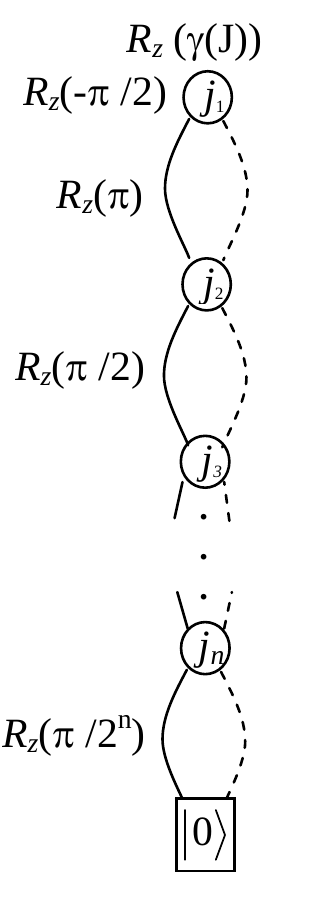}
    }
    }
    {
    \subfigure[\label{Fig:QFT2}]{
    \scalebox{0.7}{
    \includegraphics[height=4.0cm]{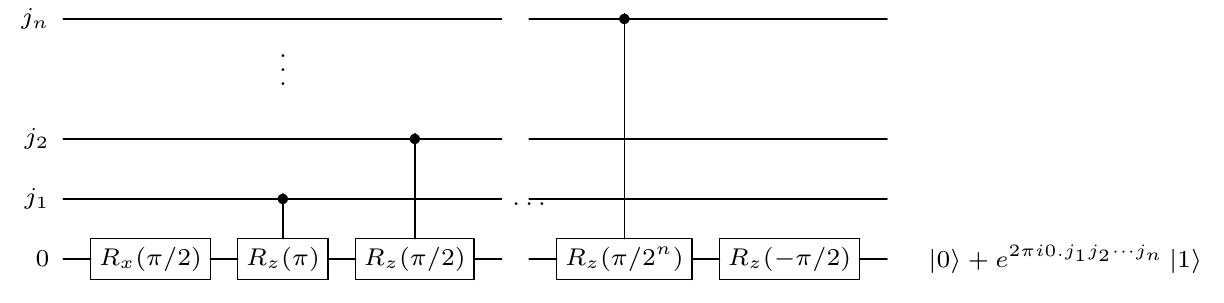}
    }
    }
    }
    \vspace{2mm}
    \caption{\label{Fig:QFT} (a) The RbDD for $R_z (\gamma (J))$ in QFT. (b) Quantum circuit for $\left| 0 \right\rangle  + e^{2\pi i0.j_1 j_2  \cdots j_n } \left| 1 \right\rangle $ in the quantum Fourier transform.}
\end{figure}

The RbDD for $R_z (\gamma (J))$ is shown in Fig. \ref{Fig:QFT1} where the root node is weighted. This RbDD corresponds to a cascade expression and there is no need to perform bi-decomposition. The quantum circuit implementing $f_n (J) = e^{i\delta (J)} R_z (\gamma (J))R_x (\theta (J))\left| 0 \right\rangle$ in shown in Fig. \ref{Fig:QFT2}.

\begin{figure}[tb]
    \scriptsize
    \centering
        \scalebox{.8}{
    \input{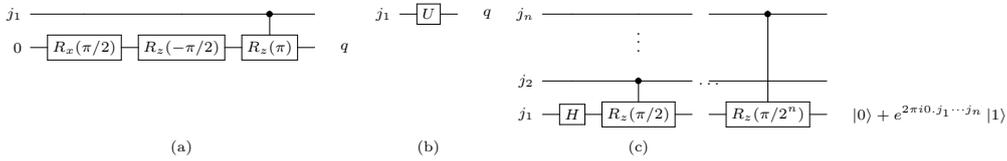}
    }
    \vspace{2mm}
    \caption{\label{Fig:QFTPart} Replacing a part of the QFT circuit (a) in Fig. \ref{Fig:QFT2} with a single qubit operator shown in (b) leads to the QFT circuit given in (c) for one output of QFT.}
\end{figure}

The single qubit operation $R_z(-\pi/2)$ can be moved between $R_x(\pi/2)$ and controlled $R_z(\pi)$ operations. Since $j_1$ is used as the controlled qubit of only one controlled rotation operation, the  sub-circuit in Fig. \ref{Fig:QFTPart}(a) can be replaced by a single qubit operator shown in Fig. \ref{Fig:QFTPart}(b). The $2 \times 2$ matrix describing $U$ consists of two columns $u_0$ and $u_1$ such that $U = [\begin{array}{*{20}c}   {u_0} & {u_1}  \\\end{array}]$ and can be obtained as follows:

$$u_0  = R_z ( - \pi /2)R_x (\pi /2)\left| 0 \right\rangle  = \frac{{e^{ - i\pi /4} }}{{\sqrt 2 }}\left[ {\begin{array}{*{20}c}
   1  \\
   1  \\
\end{array}} \right]
$$

$$
u_1  = R_z (\pi )R_z ( - \pi /2)R_x (\pi /2)\left| 0 \right\rangle  = R_z (\pi /2)R_x (\pi /2)\left| 0 \right\rangle  = \frac{{e^{ - i\pi /4} }}{{\sqrt 2 }}\left[ {\begin{array}{*{20}c}
   1  \\
   { - 1}  \\
\end{array}} \right]
$$

Hence, we have:
\[
U = \frac{{e^{ - i\pi /4} }}{{\sqrt 2 }}\left[ {\begin{array}{*{20}c}
   1 & 1  \\
   1 & { - 1}  \\
\end{array}} \right]
\]

This operator can be replaced by the Hadamard operator since the two operators differ only in a global phase. Therefore, the quantum circuit for $\left| 0 \right\rangle  + e^{2\pi i0.j_1 j_2  \cdots j_n } \left| 1 \right\rangle $  can be realized as shown in Fig. \ref{Fig:QFTPart}(c). The remaining outputs can be generated similarly. {Accordingly, the proposed method results in the same circuit structure in \cite{NeilsenChuang} with $n(n+1)/2$ total gates. This can show the efficiency of the proposed \emph{automatic} synthesis approach.} \qed

%\end{example}

\section{Conclusions and Further Discussion}\label{sec:conc}
We mainly addressed reversible logic synthesis by quantum rotation-based gates. A new canonical representation model was proposed based on binary decision diagrams. Focused on it, we developed a synthesis framework to manipulate circuits and to synthesize functions with binary variables. We also showed that the proposed approach can be extended to work with functions that generate arbitrary outputs for binary inputs.

While almost all previous synthesis methods with favorable results \cite{SaeediM2011} used CNOT, controlled-V and controlled-V$^\dagger$ gates (see Fig. \ref{Fig:CTGates-org}) as primitive gates with unit cost, we used 2-qubit controlled-rotation gates. This work can be particularly considered as a synthesis method for Boolean reversible circuits that computes a given Boolean function outside the Boolean domain with quantum gates \cite{MaslovTCAD11}. We hope this new insight opens further analysis and investigation to efficiently address quantum and reversible logic synthesis possibly beyond current achievements \cite{SaeediM2011}.

To realize a given quantum computation by fault-tolerant gates, one needs to use those gates that have direct fault-tolerant implementations \cite{NeilsenChuang}. Such realizations are only available for a few operations such as Clifford gates. To implement a wider set of gates such as the ones we used in this paper, one must apply the set of fault-tolerant gates to accurately (by approximation) implement other gates. This can be done by the Solovay-Kitaev algorithm \cite{NeilsenChuang}. Given the point that the proposed approach uses controlled rotation gates with various angles, fault-tolerant implementation of the proposed circuits can be costly. Future work should address this issue. Additionally, further progress on this path may result in new observations to restrict/ignore angles \cite{BarencoAQFT,Fowler2004} and to remove redundant gates.

\section*{Acknowledgements}
{MS thanks Alireza Shafaei for useful discussion.}
 Authors were supported in part by the Intelligence Advanced
Research Projects Activity (IARPA) via Department
of Interior National Business Center contract number
D11PC20165. The U.S. Government is authorized to reproduce
and distribute reprints for Governmental purposes
notwithstanding any copyright annotation thereon. The views
and conclusions contained herein are those of the authors and
should not be interpreted as necessarily representing the official
policies or endorsements, either expressed or implied, of
IARPA, DoI/NBC, or the U.S. Government.

\end{document}